\documentclass[11pt,a4paper]{article}
\usepackage{jheppub}

\usepackage{amsmath,amssymb,calc}
\usepackage{color}
\usepackage{graphicx}
\usepackage{ bbold }
\usepackage[utf8]{inputenc}
\makeatletter
\renewcommand\section{\@startsection {section}{1}{\z@}%
                                   {-3.5ex \@plus -1ex \@minus -.2ex}%nn
                                   {2.3ex \@plus.2ex}%
                                   {\normalfont\large\bfseries}}
\renewcommand\subsection{\@startsection{subsection}{2}{\z@}%
                                     {-3.25ex\@plus -1ex \@minus -.2ex}%
                                     {1.5ex \@plus .2ex}%
                                     {\normalfont\bfseries}}
\makeatother

\newcommand{\bea}{\begin{eqnarray}}
\newcommand{\eea}{\end{eqnarray}}
\newcommand{\be}{\begin{equation}}
\newcommand{\ee}{\end{equation}}
\newcommand{\bma}{\begin{pmatrix}}
\newcommand{\ema}{\end{pmatrix}}

\def\barray{\begin{array}}
\def\earray{\end{array}}
\def\be{\begin{equation}}
\def\ee{\end{equation}}
\def\ben{\begin{equation} \nonumber}
\def\een{\end{equation}}
\def\ban{\begin{eqnarray*}}
\def\ean{\end{eqnarray*}}
\def\ba{\begin{eqnarray}}
\def\ea{\end{eqnarray}}

\def\curv{\mathcal{R}}

\def\({\left(}
\def\){\right)}
\def\[{\left[}
\def\]{\right]}

\def\tr{{\rm Tr}}
\def\nn{\nonumber}

\def\hz{\hat{z}}

\def\f{\mathcal{F}}

\def\v{\mathcal{V}}
\def\mpsi{m_{\psi}}

\def\g{\mathcal{G}}
\def\tr{{\rm Tr}}
\def\hg{{\hat\gamma}}
\def\hy{{\hat t}}

\def\zhat{{\hat z}}
\def\phihat{{\hat \phi}}
\def\One{{\hbox{ 1\kern-.8mm l}}}

\def\ada{{\hat X}}

\def\zhat{{\hat z}}

\def\phihat{{\hat{\delta\!\phi}}}

\def\dHiggs{\xi}
\def\hvev{Z_0}
\def\mHiggs{m_{\hvev}}

\def\dA{\delta\! A}

\def\da{\Psi}

\def\axion{\mathcal{X}}
\def\daxion{\delta\!\mathcal{X}}

\definecolor{darkgreen}{cmyk}{0.85,0.2,1.00,0.2}

\def\be{\begin{equation}}
\def\ee{\end{equation}}
\def\bea{\begin{eqnarray}}
\def\eea{\end{eqnarray}}

\def\IZ{\relax\ifmmode\mathchoice
{\hbox{\cmss Z\kern-.4em Z}}{\hbox{\cmss Z\kern-.4em Z}}
{\lower.9pt\hbox{\cmsss Z\kern-.4em Z}} {\lower1.2pt\hbox{\cmsss
Z\kern-.4em Z}}\else{\cmss Z\kern-.4em Z}\fi}
\def\IR{\relax{\rm I\kern-.18em R}}
\def\One{{\hbox{ 1\kern-.8mm l}}}

\def\tr{{\rm Tr\,}}

\newlength{\bredde}
\def\slash#1{\settowidth{\bredde}{$#1$}\ifmmode\,\raisebox{.15ex}{/}
\hspace*{-\bredde} #1\else$\,\raisebox{.15ex}{/}\hspace*{-\bredde}
#1$\fi}

\newsavebox{\zzzbar}
\sbox{\zzzbar}
  {\setlength{\unitlength}{0.9em}
  \begin{picture}(0.6,0.7)
  \thinlines
  \put(0,0){\line(1,0){0.6}}
  \put(0,0.75){\line(1,0){0.575}}
  \multiput(0,0)(0.0125,0.025){30}{\rule{0.3pt}{0.3pt}}
  \multiput(0.2,0)(0.0125,0.025){30}{\rule{0.3pt}{0.3pt}}
  \put(0,0.75){\line(0,-1){0.15}}
  \put(0.015,0.75){\line(0,-1){0.1}}
  \put(0.03,0.75){\line(0,-1){0.075}}
  \put(0.045,0.75){\line(0,-1){0.05}}
  \put(0.05,0.75){\line(0,-1){0.025}}
  \put(0.6,0){\line(0,1){0.15}}
  \put(0.585,0){\line(0,1){0.1}}
  \put(0.57,0){\line(0,1){0.075}}
  \put(0.555,0){\line(0,1){0.05}}
  \put(0.55,0){\line(0,1){0.025}}
  \end{picture}}

\newcommand{\ena}{\end{eqnarray}}
\newcommand{\beqa}{\begin{eqnarray}}
\newcommand{\eeqa}{\end{eqnarray}}

\newfont{\goth}{ygoth.tfm scaled 1200}                   % gothic font (usual)
\numberwithin{equation}{section}

\renewcommand{\thesection}{\arabic{section}}

\title{Higgsed Chromo-Natural Inflation}

\author[a]{Peter Adshead,}
\author[b,c]{Emil Martinec,}
\author[a]{Evangelos I. Sfakianakis,}
\author[d]{and Mark Wyman}

\affiliation[a]{Department of Physics, University of Illinois at Urbana-Champaign, Urbana, IL 61801, USA}
\affiliation[b]{Enrico Fermi Institute,  University of Chicago, Chicago, IL 60637, USA} 
\affiliation[c]{Department of Physics,  University of Chicago, Chicago, IL 60637, USA}
\affiliation[d]{PDT Partners, 1745 Broadway, 25th Floor New York, NY 10019, USA}

\emailAdd{adshead@illinois.edu} 
\emailAdd{ejmartin@uchicago.edu} 
\emailAdd{esfaki@illinois.edu}
\emailAdd{markwy@gmail.com}

\abstract{We demonstrate that Chromo-Natural Inflation  can be made consistent with observational data if the SU(2) gauge symmetry is spontaneously broken.  Working in the Stueckelberg limit, we show that isocurvature is negligible, and the resulting adiabatic fluctuations can match current observational constraints. Observable levels of chirally-polarized gravitational radiation ($r\sim 10^{-3}$) can be produced while the evolution of all background fields is sub-Planckian. The gravitational wave spectrum is amplified via linear mixing with the gauge field fluctuations, and its amplitude is not simply set by the Hubble rate during inflation. This allows observable gravitational waves to be produced for an inflationary energy scale below the GUT scale. The tilt of the resulting gravitational wave spectrum can be either blue or red.}

\begin{document}

\maketitle
\flushbottom

\section{Introduction}

Inflation \cite{Guth:1980zm, Linde:1981mu, Albrecht:1982wi} remains a remarkably successful paradigm for describing the initial conditions of our Universe. As well as solving the flatness and horizon problems, inflation provides a mechanism for generating primordial fluctuations with the right amplitude and scale dependence to seed structure formation \cite{Mukhanov:1981xt,Chibisov:1982nx}, as well as possibly producing primordial gravitational waves \cite{Starobinsky:1979ty}. While there exist many models of inflation in the current literature, and many that fit the data well, most existing models of inflation rely on scalar fields slowly rolling on flat potentials to drive the inflationary epoch.

Chromo-Natural Inflation \cite{Adshead:2012kp} is a model for the inflationary epoch where non-Abelian gauge fields in classical, color-locked configurations generate an attractor solution which decouples the motion of the inflaton, in this case a pseudo-scalar, from the gradient flow of the potential. While Chromo-Natural Inflation can successfully generate long periods of inflation at the classical level, it fails to provide the seeds for structure formation consistent with current observations \cite{Dimastrogiovanni:2012ew,   Adshead:2013qp, Adshead:2013nka}. Furthermore, related models such as Gauge-flation \cite{Maleknejad:2011jw, Maleknejad:2011sq}, are also inconsistent with current observations once the fluctuations are taken into account \cite{Namba:2013kia}. 

In this paper we demonstrate that Chromo-Natural Inflation \cite{Adshead:2012kp} can potentially be made a viable candidate for the generation of primordial curvature perturbations by introducing an additional mass for the gauge field fluctuations via spontaneous symmetry breaking. While we dub the resulting theory Higgsed Chromo-Natural Inflation, in this work we restrict consideration to the Goldstone sector of the resulting broken gauge symmetry, and work with the action in Stueckelberg form. We  ignore the possible existence of a Higgs boson, and assume its mass  is large compared with the Hubble rate during inflation. The resulting theory can generate large levels of gravitational radiation of a single (helical) polarization only, while all background fields roll over sub-Planckian distances. Further, the amplitude of the resulting gravitational wave spectrum is not simply set by the Hubble rate, and as a result observable gravitational waves can be produced while the inflationary energy density is somewhat below the energy scale associated with grand unification. Despite consisting of several fields, isocurvature perturbations are suppressed relative to adiabatic modes. 

Admittedly, the addition of more fields is a little distasteful, an epicycle on an already speculative idea. However, it is worth pointing out that the only SU(2) gauge theory which appears to describe nature, that associated with the electroweak sector, exists in a broken phase \cite{Yang:1954ek,Anderson:1963pc,Englert:1964et,Higgs:1964ia,Guralnik:1964eu,Migdal:1966tq,Weinberg:1967tq,Glashow:1961tr}.  Further, the model we describe in this work provides an explicit counter-example to the standard inflationary lore, that the detection of tensor modes implies that inflation happened near the GUT scale and requires super-Planckian field excursions \cite{Lyth:1996im}. For a more sophisticated analyses of the field range bound see ref.\  \cite{Baumann:2011ws}, and for a more general discussion of the relation between the energy scale of inflation and the gravitational wave spectra see ref.\ \cite{Mirbabayi:2014jqa}.

Classical non-Abelian gauge fields lead to striking phenomenology in cosmological settings, most notably chiral gravitational waves \cite{Adshead:2013nka, Maleknejad:2014wsa, Obata:2014loa, Bielefeld:2014nza, Bielefeld:2015daa, Obata:2016tmo, Maleknejad:2016qjz, Caldwell:2016sut,Alexander:2016moy, Obata:2016xcr, Dimastrogiovanni:2016fuu}. These chiral gravitational waves may be responsible for the matter-antimatter asymmetry via the gravitational anomaly \cite{Alexander:2004us, Noorbala:2012fh, Maleknejad:2014wsa, Maleknejad:2016dci}. Chiral gravitational waves also arise in other models involving axially coupled gauge fields \cite{Anber:2012du, Barnaby:2012xt} and axially coupled  fermions \cite{Anber:2016yqr}. For a recent review of axion inflation see ref.\ \cite{Pajer:2013fsa} and for a recent review of gauge fields and inflation see ref.\ \cite{Maleknejad:2012fw}. 

Throughout this work, we use natural units where the speed of light and the reduced Planck constant, $c = \hbar = 1$ and the reduced Planck mass $1/\sqrt{8\pi G} = M_{\rm pl}= 1$.

%%%%%%%%%%%%%%%%%%%%%%%%%%%%%%%%%
%%%%%%%%%%%%%%%%%%%%%%%%%%%%%%%%%
%
\section{Higgsed Chromo-Natural Inflation}\label{sec:CNIbackground}
%
%%%%%%%%%%%%%%%%%%%%%%%%%%%%%%%%%
%%%%%%%%%%%%%%%%%%%%%%%%%%%%%%%%%

We consider the theory of Chromo-Natural Inflation \cite{Adshead:2012kp}, which is described by the action
\begin{align}\label{eqn:CNIaction}
\mathcal{S}_{\rm CNI} = &  \int d^{4}x\sqrt{-g}\Bigg[  \frac{1}{2}R -\frac{1}{2}(\partial\axion)^2 -V(\axion)-\frac{1}{2}\tr\[F_{\mu\nu}F^{\mu\nu}\]-\frac{\lambda}{4 f}\axion\tr\[ F\wedge F \]\Bigg] ,
\end{align}
where $\axion$ is a pseudo-scalar (axion) with associated mass scale $f$. We will assume a sinusoidal Natural inflation-like axion potential with energy scale $\mu$ and decay constant, $f \ll M_{\rm Pl}$ \cite{Freese:1990rb}:
\begin{align}\label{eqn:naturalpotential}
V(\axion)=\mu^4 \(1+\cos \(\frac{\axion}{f}\)\).
\end{align}
We emphasize that the existence of inflationary solutions is not dependent on this choice. We consider a general SU(N) gauge field, $A_\mu$, and our conventions for its covariant derivative and field strength will be the same as those outlined in \cite{Adshead:2013nka}, and detailed in appendix \ref{App:conventions}.  

The combination of the shift symmetry of the axion combined with the gauge symmetry of the vector fields strongly restricts the types of interactions that we can write down.  {One interaction omitted from consideration in~\cite{Adshead:2013nka} is the coupling of the theory to a Higgs sector which spontaneously breaks the SU(2) gauge symmetry.  In the following, we will consider this addition to the theory, and work in the limit that the mass of the Higgs is much much greater than the Hubble scale.} This means that the dynamics of fluctuations that change the Higgs mass will be irrelevant, and we can therefore ignore them. In this limit, the particular representation we choose for the Higgs does not matter, since the only relevant dynamics will be that of the Goldstone boson, whose action will be in Stueckelberg form\footnote{In Higgs representations that do not completely break the gauge symmetry, such as the adjoint representation, we must put a texture in the Higgs analogous to that of the gauge field; this is discussed in appendix \ref{app:specificmodel}.} \cite{Kunimasa:1967zza,Ruegg:2003ps}
\begin{align}\label{eqn:Higgsaction}
\mathcal{S}_{\rm H, eff} =  \int d^{4}x\sqrt{-g}\[   -g^2\hvev^2\tr\[ A_{\mu} - \frac{i}{g}U^{-1}\partial_\mu U\]^2\]
\end{align}
where 
\begin{align}
U = \exp\[ig \dHiggs\],\quad \xi  =  \dHiggs^a J_a,
\end{align}
and $\dHiggs^a$ are the Goldstone modes corresponding to fluctuations of the Higgs along its vacuum manifold, and $J^a$ are the generators of the gauge symmetry. Under an infinitesimal gauge transformation, 
\begin{align}
\dHiggs^a \to \dHiggs^a-\alpha^a, \quad 
A^a_\mu \to A^a_\mu+ \partial_\mu \alpha^a + g\epsilon^{a}{}_{bc}A^b_{\mu}\alpha^c,
\end{align}
and thus eq.\ \eqref{eqn:Higgsaction} is gauge invariant.

While the main part of this work will restrict to the consideration of the dynamics of the model in the Stueckelberg limit, in appendix \ref{app:specificmodel} we describe a specific realization away from this limit and demonstrate that the SU(2) gauge symmetry can be dynamically broken in such a way as to generically preserve the background SO(3) symmetry of spacetime.

%%%%%%%%%%%%%%%%%%%%%%%%%%%%%%%%%%
%
\subsection{Background solutions}\label{sec:backgroundeqns}
%
%%%%%%%%%%%%%%%%%%%%%%%%%%%%%%%%%%

We find inflationary trajectories in the above action by considering the axion in a classical, homogeneous configuration  $\axion =  \axion(t)$  and the gauge fields in the classical configuration
\begin{align}\label{eqn:gaugevev}
A_{0} = & 0, \quad A_{i} =  \phi\delta^{a}_{i} J_{a} = a\psi \delta^{a}_{i} J_{a},
\end{align}
where $J_a$ is a generator of SU(2) satisfying the commutation relations, and normalization
\begin{align}
\[J_a, J_b\] = if_{abc}J_c, \quad {\rm Tr}\[J_a J_b\] = \frac{\delta_{ab}}{2},
\end{align}
and $f_{abc}$ are the structure functions of SU(2). Note that for SU(2), $f_{ijk} = \epsilon_{ijk}$, where $\epsilon_{ijk}$ is the completely antisymmetric tensor in three dimensions.

On the background field configuration in eq.\ (\ref{eqn:gaugevev}), the field strength tensor components are,
\begin{align}
F_{0i} = & \partial_{\tau}\phi \delta^{a}{}_{i}J^a,\quad
F_{ij}  =  g \phi^2 f^{a}_{ij} J^a,
\end{align}
where we work with conformal time, $\tau$.
 
For these degrees of freedom, the mini-superspace action takes the form
\begin{align}
\mathcal{L} =  a^{3}N \Bigg[-3  \frac{\dot a^2}{N^2} + \frac{a^2}{2 N^2} \dot \axion^2 - V(\axion) + \frac{3}{2}\frac{\dot \phi^2}{N^2}  - \frac{3}{2}g^{2}\frac{\phi^4}{a^4}-\frac{3}{2}g^2\hvev^2 \frac{\phi^2}{a^2} \Bigg]- 3\frac{\lambda}{f}g\axion \dot\phi \phi^2,
\end{align}
where here and in what follows an overdot  represents a derivative with respect to cosmic time, and the lapse, $N = a$ on the background solution. This action leads to the equations of motion for the axion $\axion$ and gauge field vacuum expectation value (VEV) $\phi$:
\begin{align}
\ddot\axion+3 H \dot\axion+ V'(\axion) = &  -\frac{1}{a^3}\frac{\lambda}{f}g\partial_{t}\(\phi^3\),\\
\frac{\ddot{\phi}}{a}+H\frac{\dot{\phi}}{a} + 2 g^2\frac{\phi^3}{a^3}+g^2\hvev^2 \frac{\phi}{a} =  & \frac{\lambda}{f}g\dot\axion \frac{\phi^2}{a^2}.
\end{align}
The equations of motion for the metric are the Friedmann constraint
\begin{align}
3 H^{2} =   \frac{1}{2} \dot \axion^2 + V(\axion) + & \frac{3}{2}\(\frac{\dot\phi^2}{a^2} + g^{2}\frac{\phi^4}{a^4}+g^2\hvev^2\frac{\phi^2}{a^2}\) ,
\end{align}
and the equation of motion for the scale factor
\begin{align}\label{eqn:metriceom}
\dot{H} = -\frac{\dot\axion^2}{ 2} - \frac{\dot{\phi}^2}{ a^2} - g^2\frac{\phi^4}{a^4} - \frac{1}{2}g^2\hvev^2\frac{\phi^2}{a^2} .
\end{align}
In \cite{Adshead:2012kp} we showed that, in the absence of the Higgs terms, this model inflates. In the limit of large $\lambda$, terms linear in time derivatives dominate the dynamics, and slow-roll is facilitated by a magnetic-drift type force mediated by the Chern-Simons interaction \cite{Martinec:2012bv}. The addition of the Higgs only slightly modifies the dynamics, and it is easily seen that similar magnetic drift type trajectories are also present in this theory. In the large drift force limit ($\lambda \gg 1$), the slow-roll equations for this model are very well approximated by
\begin{align}
\dot \psi & = - H \psi - \frac{f }{3 g \lambda} \frac{V_{,\axion}}{\psi^{2}}  
~,\quad \label{slowrollpsi}\\
\frac{\lambda}{f} \dot \axion %&
 & = 2 g \psi +\frac{g\hvev^2}{\psi}+  \frac{2H^2}{g \psi}  ~\label{slowrollX}\ ,
\end{align}
where $\psi = \phi /a$. To a good approximation, $\psi \approx {\rm const.} $, and eq.\ (\ref{slowrollpsi}) is solved by
\begin{align}
\psi = \(\frac{-f V_{,\axion}}{3 g H \lambda}\)^{1/3},
\label{eq:psimin}
\end{align}
which has been used to simplify eq.\ (\ref{slowrollX}). It will also prove useful to introduce the dimensionless mass parameters
\begin{align}
m_{\psi} = \frac{g\psi}{H}, \quad {\rm and} \quad M_{\hvev} = \frac{g\hvev}{H},
\end{align}
which characterize the various contributions of mass to the gauge field fluctuations in units of the Hubble scale. In terms of these quantities, eq.\ (\ref{eqn:metriceom}) can be written
\begin{align}
\epsilon_H \equiv -\frac{\dot H}{H^2}= &~ \frac{\dot\axion^2}{ 2H^2} + \frac{\dot\psi^2}{H^2} + 2\frac{\dot\psi}{H} \psi+\(1+\mpsi^2+\frac{M_{\hvev}^2}{2}\)\psi^2\\
= &~\epsilon_{\axion} + \(1+ \eta_{\psi}^2+2\eta_{\psi}+\mpsi^2+\frac{M_{\hvev}^2}{2}\) \epsilon_{\psi} 
\end{align}
where we introduce the slow roll parameters,
\begin{align}
\epsilon_{\axion} = \frac{\dot{\axion}^2}{2H^2}, \quad \eta_{\axion} = \frac{\ddot{\axion}}{\dot{\axion}H}, \quad \epsilon_{\psi} = \psi^2, \quad \eta_{\psi} = \frac{\dot{\psi}}{H\psi}.
\end{align}
The addition of the Higgs VEV does not prevent the existence of inflationary background solutions and the condition for inflation remains
%\begin{align}
$\epsilon_H < 1$. 
%\end{align}
Thus, in the limit where the gauge field is approximately static, $\eta_{\psi}\ll 1$, the gauge field  VEV  $\epsilon_{\psi} = \psi^2$ limits how large the Higgs VEV can be.
%%%%%%%%%%%%%%%%%%%%%%%%%%%%%%%%%%%%
%%%%%%%%%%%%%%%%%%%%%%%%%%%%%%%%%%%%

\subsection{Background parameter scan}\label{sec:backgroundparams}

\begin{figure}
\includegraphics[width=\textwidth]{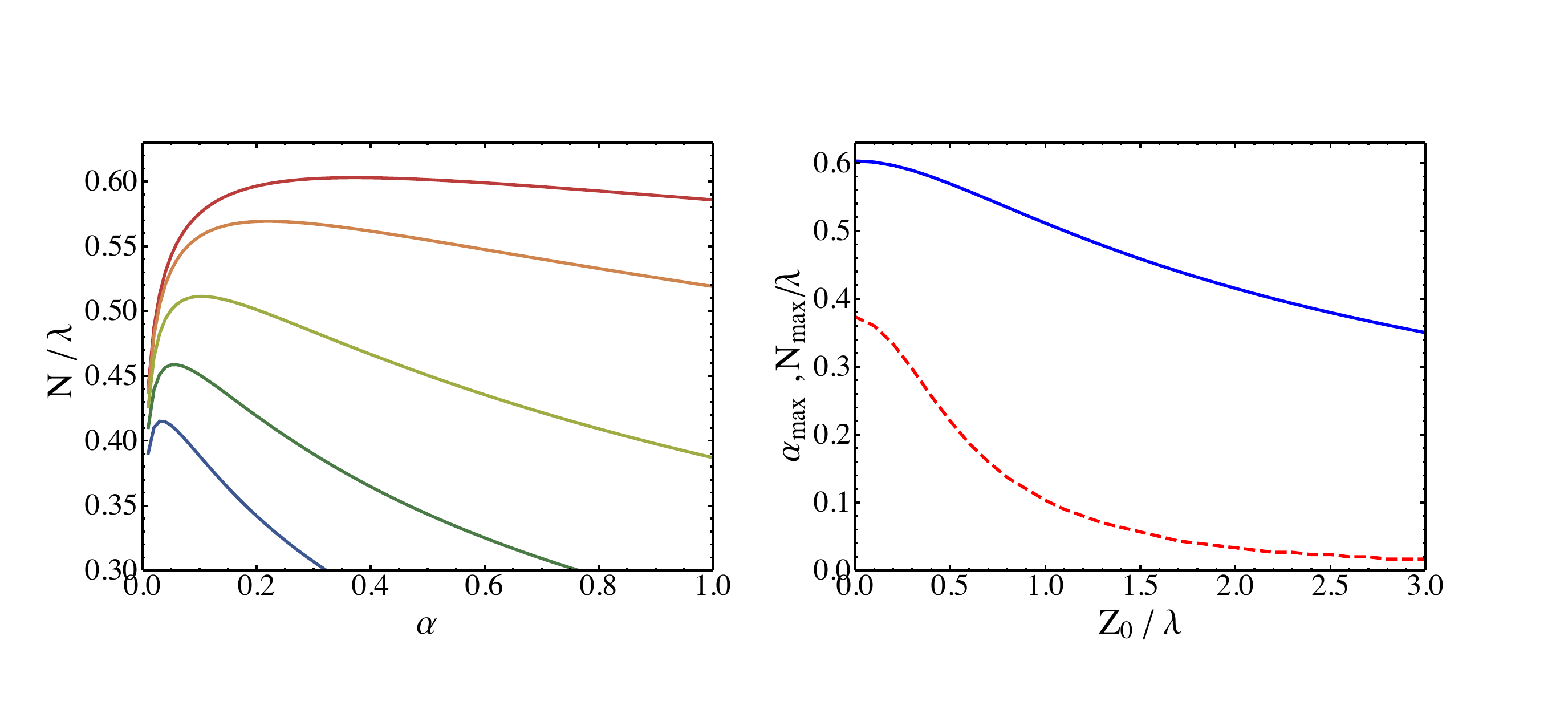}
\caption{Left: The number of e-folds as a function of the parameter $\alpha  \equiv g^2 /( \lambda \mu^4)$ for $z_0 = 0, 0.5, 1, 1.5, 2$, color-coded in a rainbow scale with red corresponding to  $Z_0=0$. Right: The maximum number of e-folds (blue) and the value of $\alpha$, where the maximum occurs (red dashed). }
\label{fig:Nmaxalphamax}
\end{figure}

Before proceeding to the analysis of the fluctuations in this model, it is worth examining the parameter dependence of the total number of e-folds of inflation.
We start by substituting the value of $\psi$ given in eq.\ \eqref{eq:psimin} into eq.\ \eqref{slowrollX}, leading to
\begin{align}
{\lambda \over f} \dot \axion = \frac{2 \sqrt[3]{3} g \left(\mu ^4 \sin \left(\frac{\chi }{f}\right)\right)^{2/3}+3 g^{5/3} Z_0^2 (H \lambda )^{2/3}+6 \sqrt[3]{\frac{H^8 \lambda ^2}{g}}}{3^{2/3} \sqrt[3]{g H \lambda  \mu ^4 \sin \left(\frac{\chi }{f}\right)}}
\end{align}
We now rescale the axion field amplitude as $x = \axion / f$ and use the number of $e$-folds $N$ as our time variable ($dN = H dt$). We can integrate the resulting expression, taking inflation to start at an axion value $x = \axion_0/f $ and end at $x =  \pi$ obtaining
\begin{align}
N(\axion_0) = \int_{\axion_0 \over f} ^\pi 
  \frac{\sqrt[3]{3} \mu ^4 \sqrt[3]{g^2 \lambda ^4 \sin (x) (\cos (x)+1)^2}}{3 Z_0^2 \sqrt[3]{g^6 \lambda ^2 \mu ^4 (\cos (x)+1)}+2 \left(\sqrt[3]{{\lambda ^2 \mu ^{16} (\cos (x)+1)^4}}+3^{2/3}  \left(g^2 \mu ^4 \sin (x)\right)^{2/3}\right)}  
dx
\label{eq:Nint}
\end{align}
Reference \cite{Adshead:2012qe} demonstrated that the maximum number of e-folds in the case of Chromo-Natural Inflation ($Z_0=0$) occurs for a specific relation of parameters, namely $g^2 /\lambda \simeq \mu^4 /3$. The present case is more complicated, since there is one further parameter to consider, $Z_0$. We start by performing the substitution $g^2 = \alpha \lambda \mu^4 $, where $\alpha$ is a numerical factor. This makes the integrand independent of $\mu$. For $Z_0=0$ the scaling $N\propto \lambda$ emerges for  $g^2 = \alpha \lambda \mu^4 $. This scaling is broken for $Z_0\ne 0$ but can be formally recovered  by defining $z_0 = Z_0 \sqrt \lambda$, which simplifies the analysis somewhat. For $Z_0=0$ the maximum number of $e$-folds is $N \simeq 0.6 \lambda$ and occurs for $\alpha \simeq 0.36$. As one increases the rescaled Higgs  VEV  $z_0$, the decrease in $\alpha$ is much more dramatic than the decrease in $N_{\rm max} / \lambda$. Furthermore, as we increase $z_0$ the number of $e$-folds falls much more quickly as we move away from its maximum, as a function of $\alpha$, as shown in figure \ref{fig:Nmaxalphamax}. 

\begin{figure}
\includegraphics[width=\textwidth]{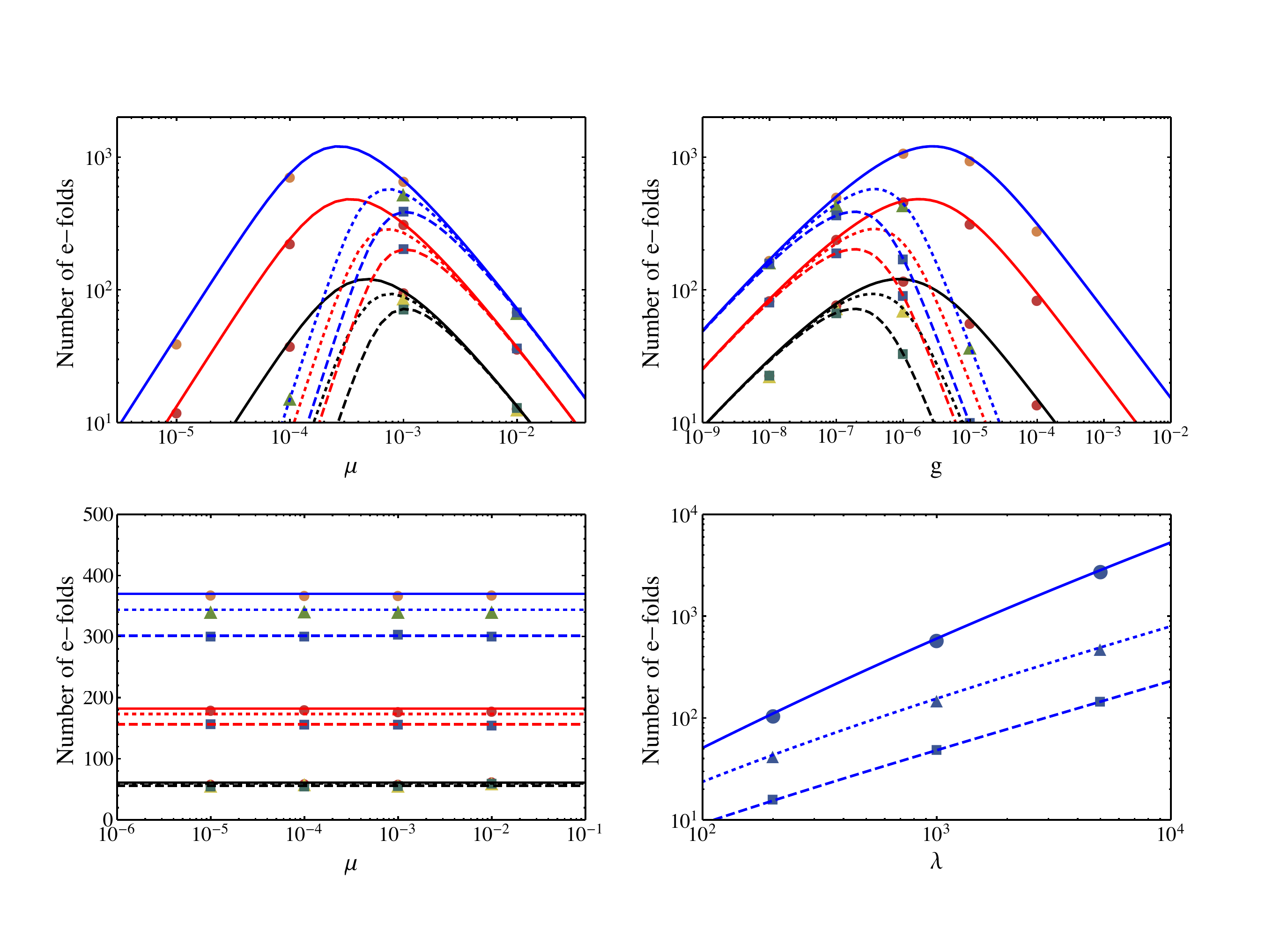}
\caption{We present the total number of e-folds of inflation for various parameter choices. The solid/dashed/dotted lines correspond to the analytical values for $Z_0=0,0.1,0.2$ respectively and the dots/triangles/squares show the results of numerical simulations.
Upper left: The total amount of inflation varies as we vary the energy scale of the axion potential $\mu$. Upper right: The total amount of inflation varies as we vary the gauge field coupling strength $g$. Lower left: The number of $e$-foldings of inflation is kept constant if the gauge field strength and the axion energy scale are varied in a way that keeps a constant ratio $ g/\mu^2=0.5$. 
We also show the effect of setting $\lambda=200, 800, 2000$ (black, red, blue respectively). Lower right: 
The total amount of inflation scales almost linearly with $\lambda$ when keeping the remaining parameters fixed. Unless otherwise noted, the remaining parameters are fixed at the values $\{\mu, f ,  g , \lambda\} = \{ 3.16 \times 10^{-4}, 0.01, 2.0 \times 10^{-6}, 200 \}$, as in \cite{Adshead:2012qe}. }
\label{fig:backgroundparameterscan}
\end{figure}

The result of varying the potential parameters on the duration of inflation is shown in figure \ref{fig:backgroundparameterscan}.  We use the approximate expression of eq.\ \eqref{eq:Nint}, setting $\axion_0 =0$, as well as a numerical evaluation of the full second-order system of equations for $\axion(t)$, $\phi(t)$ and $H(t)$. It is worth noting that as $\axion_0 \to 0$ it becomes numerically more difficult to approach the axion-gauge inflationary attractor. Choosing  $\axion_0 \simeq 0.1 f$ gives numerically well-behaved results and the difference between $N(\axion_0=0)$ and $N(\axion_0=0.1f)$ is at the level of a few percent. In general, increasing $Z_0$ reduces the number of e-folds of inflation, unless  $g$ is very small, or $\mu$ is  very large, both leading to the combination $g/\mu^2$ being small. Furthermore, the analytically and numerically derived values of $N$ are in excellent agreement for a wide range of parameters.

%%%%%%%%%%%%%%%%%%%%%%%%%%%%%%%%%%%%
%%%%%%%%%%%%%%%%%%%%%%%%%%%%%%%%%%%%
%
\section{The quadratic fluctuation action}\label{sec:quadaction}
%
%%%%%%%%%%%%%%%%%%%%%%%%%%%%%%%%%%%%
%%%%%%%%%%%%%%%%%%%%%%%%%%%%%%%%%%%%

In order to find the equations of motion for small fluctuations about the above background solutions, we compute the action to quadratic order in fluctuations. The variation of this quadratic action will yield the linear equations of motion.\footnote{In practice, we make use of {\sc Mathematica} to obtain the action for the quadratic fluctuations and the resulting equations of motion.}

We work with the metric in ADM form \cite{Arnowitt:1962hi},
\begin{align}\label{eqn:adm}
ds^2 = -N^2 d\tau^2 +  \tilde{h}_{ij}(dx^i+N^i d\tau)(dx^j+N^j d\tau),
\end{align}
where $N$ is the lapse function, $N^i$ is the shift vector, and $\tilde{h}_{ij}$ is the metric on the spatial hypersurface. In our conventions, the background Friedmann-Robertson-Walker (FRW) metric in conformal coordinates corresponds to $N = a$ and $N^i = 0$. The diffeomorphism invariance of general relativity allows us to choose our coordinates so that spatial hypersurfaces are Ricci flat. This gauge choice completely fixes the coordinates. We further write\footnote{Our summation convention is as follows. Repeated lower Roman indices and all gauge field indices are summed with the Kronecker delta. }
\begin{align}\label{eqn:spatialmet2}
\tilde{h}_{ij} = a^2\[e^{\gamma}\]_{ij} = a^2\left[\delta_{ij} + \gamma_{ij}+\frac{1}{2!}\gamma_{ik}\gamma_{kj}+\ldots\right],
\end{align}
so that $\gamma$ is a spin-2 mode of the metric.

It was demonstrated in ref.\ \cite{Dimastrogiovanni:2012ew} that the contributions of the lapse and shift constraints have no effect on the equations of motion until after horizon crossing, where the fluctuations have frozen out. Further,  in \cite{Adshead:2013nka} we demonstrated that the contributions to the action due to integrating out the lapse and the shift (i.e. solving the Einstein constraints) are suppressed relative to the contributions to the action from the non-gravitational terms by small background quantities such as $\dot\axion$ and $\psi$.  The same is true in the case at hand, and we ignore contributions to the action from the gravitational constraints as well as contributions due to the evolution of the background. 

Defining the fluctuations of the gauge field by
\begin{align}
\dA_\mu = \Psi_\mu
\end{align}
and using the background field configuration of eq.\ (\ref{eqn:gaugevev}), the quadratic Yang-Mills Lagrangian density can be written,
\begin{align}\label{eqn:actYM}\nn
\delta^2\mathcal{L}_{\rm YM} = &\tr\[(\partial_{i}\da_{0}-i g\phi\[J_i, \da_0\])^2\] -4ig\partial_{\tau}\phi\tr\[\da_0\[\da_i, J_i\]\]
%
%\\ \nn&
-2\tr\[\da_{0}\partial_{\tau}(\partial_{i}\da_{i}-ig\phi \[J_{i}, \da_i\])\]
\\ \nn& +\tr\[\partial_{\tau}\da_i\partial_{\tau}\da_i\] -\tr\[\partial_j \da_{i}\partial_{j}\da_{i}-\partial_i \da_{j}\partial_{j}\da_{i}\]+2g\phi\epsilon_{ijk}\tr\[\partial_i \da_{j}\Omega_k \] \\ &-g^2\phi^2\tr\[(\Omega_k-\da_{k})\Omega_{k}\],
\end{align}
where we have defined
\begin{align}\label{eqn:omegadef}
\Omega_i = i\epsilon_{ijk}\[J_j,\da_k\].
\end{align}
Similarly, the quadratic order Chern-Simons Lagrangian density can be written,
\begin{align}\label{eqn:actCS}\nn
\delta^2\mathcal{L}_{\rm CS} = &  2g\phi^2\frac{\lambda}{f}\daxion\tr\[\partial_i \da_0 J_i\] -\frac{\lambda}{f}\partial_{\tau}\axion\tr\[  g\phi \da_i \Omega_{i}-\epsilon_{ijk}\da_i\partial_{j}\da_{k}\]  + 2g\phi^2\frac{\lambda}{f}\partial_{\tau}\daxion\tr\[ \da_i J_i\]\\ & - 2\frac{\lambda}{f}\epsilon_{ijk}\partial_{\tau}\phi
\daxion \tr\[J_i \partial_j \da_k\].
\end{align}
The axion contribution to the quadratic Lagrangian density is
\begin{align}\label{eqn:actaxion}
\delta^2\mathcal{L}_{\axion} = &   \frac{1}{2}a^2(\partial_{\tau} \daxion )^2- \frac{1}{2}a^2(\partial_{i} \daxion )^2 - a^4\frac{1}{2} \frac{d^2V}{d\axion^2}\daxion^2.
\end{align}
The quadratic Lagrangian density for the transverse-traceless components of the metric, and their interactions with the gauge field fluctuations is given by
\begin{align}\nn\label{eqn:actspin2}
\delta^2\mathcal{L}_{\gamma} = & \frac{a^2}{8} \((\partial_{\tau}\gamma)^{2}-(\partial_i \gamma)^{2}+2\(\dot\phi^2-g^{2}\frac{\phi^4}{a^2}\)\gamma^2\)\\  & -a^2 \(\frac{\dot{\phi}}{a} \partial_{\tau}\da_{jl} -g\frac{\phi^2}{a^2}(2\epsilon^{a}_{ij}\partial_{[i}\da^a_{l]}+g\phi \da_{jl}) \)\gamma_{jl},
\end{align}
where $\gamma^2 = \gamma_{ij}\gamma_{ij}$. These four contributions make up the action for the original theory of Chromo-Natural Inflation and the above was presented in \cite{Adshead:2013nka}. The key addition that we are introducing in this work is the interaction of the gauge field with a Higgs sector. As we have mentioned above, we are assuming that fluctuations that change the mass of the Higgs will be irrelevant, since these are much more massive than the Hubble scale. We thus restrict to the Goldstone bosons which fluctuate along the vacuum manifold, these contribute at quadratic order in fluctuations via
\begin{align}\nn
\delta^2\mathcal{L}_{\rm Higgs} = & a^4\Bigg[-\frac{g^2\hvev^2}{2}g^{\mu\nu}\(\partial_{\mu}\dHiggs^a + \da^a_\mu\)\(\partial_{\nu}\dHiggs^a + \da^a_\nu\)+\frac{g^2\hvev^2{g\psi}}{a}\epsilon_{bic}\dHiggs^b\partial_{i}\dHiggs^c \\ &\hskip 2cm 
 -\frac{g^2\psi^2\hvev^2}{4} \gamma^2+ g^2 \frac{\hvev^2\psi }{a}\gamma_{ij}\da_{ij}\Bigg].
\end{align}
The addition of a Higgs sector thus yields an additional mass term for the gauge field fluctuations. Note, however, retaining gauge-invariance requires us to also add the Goldstone modes $\dHiggs$.

Following \cite{Adshead:2013nka}, we work with a 2-dimensional representation of the gauge field and decompose the fluctuations in the gauge field as 
\begin{align}\label{eqn:fielddecomp}
 \da^a_i = \(t^a_i + \epsilon^a_{ij} \chi^j+\delta^a_i \delta\phi\)J_a.
\end{align}
We also work with explicit components of the fields. Choosing the wavenumber along the $x^3$ direction, the gauge-field modes eq.\ (\ref{eqn:fielddecomp}) then have a scalar-vector-tensor (SVT) decomposition in which 
\begin{align}
t^\pm =\frac{1}{\sqrt{2}}\( \frac12(t_{11}-t_{22})\pm i t_{12}  \)
\end{align}
forms the two helicities of a transverse traceless tensor, 
\begin{align}
v^{\pm} = \frac{1}{\sqrt{2}}\(t_{3,1}\pm i t_{3,2}\), \quad u^{\pm} = \frac{1}{\sqrt{2}}\(\chi_{1} \pm i\chi_{2}\)
\end{align}
are helicity states of transverse vectors, and 
\begin{align}
z\equiv\frac16(2t_{33}-t_{11}-t_{22}),
\end{align}
is a scalar along with  $ \chi_3$, and $\delta\!\phi$. Rotational invariance ensures that the particular choice of direction is irrelevant, and thus we drop the `3' subscript on the momenta. Additionally, the Goldstone modes can be similarly decomposed into a scalar mode $\dHiggs \equiv \dHiggs^3$ and two vector modes
\begin{align}
\xi^{\pm} \propto \xi^1\pm i\xi^2.
\end{align}

The SU(2) gauge invariance of the action allows us to fix a gauge for the gauge field fluctuations and eliminate three of the degrees of freedom in the gauge sector. Observable quantities, such as the components of the energy-momentum tensor, are by definition gauge invariant. This means as long as the gauge is completely fixed, physical quantities will not  be dependent on the particular choice of gauge.  In this work, we will work in a non-Abelian generalization of the Coulomb condition \cite{Adshead:2013nka}
\begin{align}\label{eqn:NAGcond}
\bar{D}_i \da_i = \partial_{i}\da_i  - i g \phi \[J_i, \da_i\] = 0,
\end{align}
dubbed non-Abelian Coulomb gauge.  A nice property of this gauge choice is that it eliminates time derivatives of the gauge field from solution of the Gauss law constraint. There are of course many other choices on might make. The works of \cite{Dimastrogiovanni:2012st, Dimastrogiovanni:2012ew, Namba:2013kia} chose to work in a gauge where $\da^a_i$ was symmetric, this is equivalent to setting to zero the field $\chi^i$. In analogy with particle physics, one may choose to work in unitary gauge, where the Higgs fluctuations are chosen to be zero $\dHiggs^a = 0$. Alternatively, one may work with combinations of the field fluctuations which are invariant under SU(2) gauge transformations \cite{Maleknejad:2011jw,Maleknejad:2011sq}.

In terms of the field decomposiiton in the two-dimensional representation, the non-Abelian Coulomb gauge condition, eq.\ (\ref{eqn:NAGcond}), additionally imposes a relationship between the degrees of freedom 
\begin{align}\label{eqn:nacoulombgauge}
\partial_i(t^a_i+\epsilon^a_{ij}\chi_j+\delta^a_{i}\delta\!\phi)= 2 g \phi \chi^a.
\end{align}
In terms of the above fields, the gauge condition as written in eq.\ (\ref{eqn:nacoulombgauge}) becomes, 
\begin{align}\label{eqn:vecgaugecon}
-ik(v_{\pm}  \pm iu_{\pm} )= & 2 g \phi u_{\pm},\\
-i k (2z +\delta\phi) = & 2 g \phi \chi_3.\label{eqn:scalargaugecond}
\end{align}
These three conditions remove three degrees of freedom. The Gauss law constraint, or the equation of motion for the non dynamical temporal component of the gauge field $A_0$, removes three further degrees of freedom leaving the six physical propagating degrees of freedom of the gauge theory. 

Ignoring the gravitons for a moment, on the above decomposition, the action reads
\begin{align}\label{eqn:action}
\delta^2 \mathcal{L} = &   - \frac{a^2}{2}(\partial_\tau \daxion )^2 - a^4\frac{1}{2}  \frac{d^2V}{d\axion^2}\daxion^2
\\\nn  &  + a^2 g^2\hvev^2 \da^d_0 \partial_{\tau }\dHiggs^d_a+\frac{a^2 g^2 \hvev^2}{2}\partial_{\tau}\dHiggs^a\partial_{\tau}\dHiggs^a - \frac{a^2 g^2 \hvev^2}{2}\partial_{i}\dHiggs^a \partial_{i}\dHiggs^a  -a^2 g^2 \hvev^2\(t^a_i+\epsilon^a_{ij}\chi^j+\delta^a_i \delta\phi\)\partial_{i}\dHiggs^a
\\\nn 
&+g^2\hvev^2 a^2{g\psi}\epsilon_{bic}\dHiggs^b\partial_{i}\dHiggs^c +\frac{1}{2}\da_0^a(-\partial^2 +2g^2\phi^2+g^2 a^2 \hvev^2)\da_0^a+g\phi \epsilon^{i}_{ba}\da^b_0 \partial_i \da^a_0\\ \nn& +\da_0^a ( \partial_{\tau}\partial_i(t^a_i + \epsilon^a_{ij}\chi^j+\delta^a_i\delta\phi)-2 \partial_{\tau}(g\phi\chi^a)+4 g \partial_{\tau}\phi  \chi^a  - g\phi^2\frac{\lambda}{ f}\partial_{a}\daxion  )\\\nn & 
+\frac{1}{2}\partial_{\tau}t^a_i\partial_{\tau}t^a_i+\partial_\tau \chi_i \partial_\tau \chi_i +\frac{3}{2}\partial_\tau \delta\phi \partial_\tau \delta\phi -  \frac{1}{2}\partial_{j}t^a_i\partial_{j}t^a_i-\partial_j\chi_i \partial_j \chi_i -\frac{3}{2}\partial_j\delta\phi \partial_j \delta\phi \\\nn
&+\frac{1}{2}\partial_i(t^a_i+\epsilon^a_{ij}\chi^j+\delta^a_{i}\delta\phi)\partial_k(t^a_k+\epsilon^a_{kj}\chi^j+\delta^a_{k}\delta\phi)+g\phi \big(\epsilon^{ijk}\partial_i t^a_j t^a_k\; - \epsilon^{ijk}\partial_i \chi^j\chi^k +6\partial_i \chi^i\delta\phi\big)\\\nn & 
-\frac{1}{2}g^2\phi^2(t^a_it^a_i + 2 \chi_i \chi_i+12\delta \phi^2)-\frac{1}{2} g\phi\(\frac{\lambda}{f}\partial_{\tau}\axion-g\phi\) \(t^a_i t^a_i - 2\chi_i \chi_i - 6\delta\phi^2\)\\\nn
 & - \frac{g^ 2 a^2\hvev^2}{2} (t^a_i t^a_i+2\chi^i\chi^i+3\delta\phi^2) \\\nn
& +\frac{\lambda}{2f}\partial_{\tau}\axion\big(\epsilon^{ijk}\partial_it^a_{j}t^a_k- 2\partial_it^i_{j}\chi^j  -\epsilon_{ijk}\partial_i\chi^k\chi^j -4 \delta\phi \partial_i\chi^i \big)+3 g\phi^2\frac{\lambda}{f}\partial_{\tau}\daxion  \delta\phi+ 2\frac{\lambda}{f}\partial_{\tau}\phi\daxion \partial_{j}\chi^j,
\end{align}
where we have not fixed the gauge for the SU(2). The coordinates have been chosen according to eq.\ (\ref{eqn:spatialmet2}) and as described above, we have ignored the contributions due to the gravitational constraint equations.

\section{Scalars}\label{sec:scalarmodes}

We begin by examining the behavior of the scalar modes. The dynamics of these modes follows from the scalar parts of the action, which reads
\begin{align}\nn\label{eqn:action}
\delta^2 \mathcal{L} = &    \frac{a^2}{2}\partial_{\tau} \daxion \partial_{\tau} \bar{\daxion} - \frac{a^2}{2}k^2\daxion\bar{\daxion} - a^4\frac{1}{2}  \frac{d^2V}{d\axion^2}\daxion^2
\\\nn 
& -\frac{1}{2}(k^2 +2g^2\phi^2+g^2 a^2 \hvev^2)^{-1} |4 g \partial_{\tau}\phi  \chi_3+a^2g^2 \hvev^2  \partial_{\tau }\dHiggs + ik g\phi^2\frac{\lambda}{ f}\daxion  |^2\\\nn & 
+3\partial_{\tau}z\partial_{\tau}\bar{z}+\partial_\tau \chi_3 \partial_\tau \chi_3 +\frac{3}{2}\partial_\tau \delta\phi \partial_\tau \delta\phi -  3k^2z\bar{z}-k^2\chi_3  \bar{\chi}_3 -\frac{3}{2}k^2 \delta\phi \bar{ \delta\phi }\\ \nn
&+\frac{a^2g^2 \hvev^2}{2}\partial_{\tau}\dHiggs \partial_{\tau}\bar{\dHiggs} -\frac{a^2g^2 \hvev^2}{2}k^2 \dHiggs \bar{\dHiggs}-g \frac{a^2g^2 \hvev^2}{2}\(ik\(2 z+\delta\phi\) \bar{\dHiggs} +c.c.\) \\\nn
 & -\frac{g^2a^2\hvev^2}{2} (6z\bar{z}+2\chi_3\bar{\chi}_3+3\delta\phi\bar{\delta\phi}) +3g\phi ik (\bar{\chi}_3\delta\phi - {\chi}_3\bar{\delta\phi })\\\nn & 
-9g^2\phi^2\delta \phi \bar{\delta \phi}-\frac{1}{2} g\phi \frac{\lambda}{f}\partial_{\tau}\axion \(6 z\bar{z} - 2\chi_3 \bar{\chi}_3 - 6\delta\phi\bar{\delta\phi}\)\\\nn
& +ik\frac{\lambda}{f}\partial_{\tau}\axion\big( z\bar{\chi}_3 - \bar{z}\chi_3+\bar{\delta\phi}\chi_3 - \bar{\chi}_3\delta\phi\big)+\frac{3}{2} g\phi^2\frac{\lambda}{f}\(\partial_{\tau}\bar{\daxion}  \delta\phi +\partial_{\tau}\daxion  \bar{\delta\phi} \)\\ & + ik\frac{\lambda}{f}\partial_{\tau}\phi(\daxion \bar{\chi_3} - \bar{\daxion}\chi_3).
\end{align}
Note that, in this expression  we have integrated out the Gauss law constraint associated with the temporal components of the gauge field, the solution of which is
\begin{align}\label{eqn:scalargauss}
\da^3_0 = -\frac{4 g \partial_{\tau}\phi  \chi_3+a^2g^2 \hvev^2  \partial_{\tau }\dHiggs +ik g\phi^2\frac{\lambda}{ f}\daxion}{(k^2 +2g^2\phi^2+g^2 a^2 \hvev^2)}  
\end{align}
We have also made use of the gauge condition eq.\ (\ref{eqn:NAGcond}) to simplify several terms, however, we have not yet completely imposed the gauge condition. It remains to  eliminated one of the degrees of freedom in order to obtain an action that contains only dynamical degrees of freedom. In this work we choose to eliminate $\chi_3$ using eq.\ (\ref{eqn:NAGcond}), although there is nothing special about this choice. The elimination of $\chi_3$ kinetically couples the fields, and we shift and rescale the fields
\begin{align}
\phihat = &\sqrt{2}\sqrt{2+\frac{k^2}{g^2 \phi^2}}\(\frac{\delta\phi}{2} + z\),\quad
\zhat =   \sqrt{2}(z - \delta\phi), \\ \ada & = a\, \daxion, \quad \hat{H} = g\hvev \sqrt{\frac{2g^2\phi^2+k^2}{g^2a^2\hvev^2+8g^2\phi^2+4k^2}}i a\xi
\end{align}
which puts the action in canonical form. Writing,
\begin{align}
\Delta = (\phihat, \zhat, \ada, \hat{H}),
\end{align}
and after integration by parts and the neglect of a boundary term the action can be put in the form
\begin{align}
\mathcal{S} = \frac{1}{2}\int d^3 x d\tau \[ \Delta^{\dagger}{}' T \Delta' + \Delta^{\dagger}K \Delta'- \Delta^{\dagger}{}'K \Delta - \Delta^{\dagger} \Omega^2  \Delta \], \quad T = \mathbb{1}.
\end{align}
The matrix $K$ is anti-Hermitian and $\Omega^2$ is symmetric. Their exact forms are not particularly illuminating, and so we omit reproducing them here, however, they are reproduced in appendix \ref{app:scalargore}.

\subsection{Quantization and initial conditions}
\label{initialconditions}
\begin{figure}[t!]
\centering
\includegraphics[width=\textwidth]{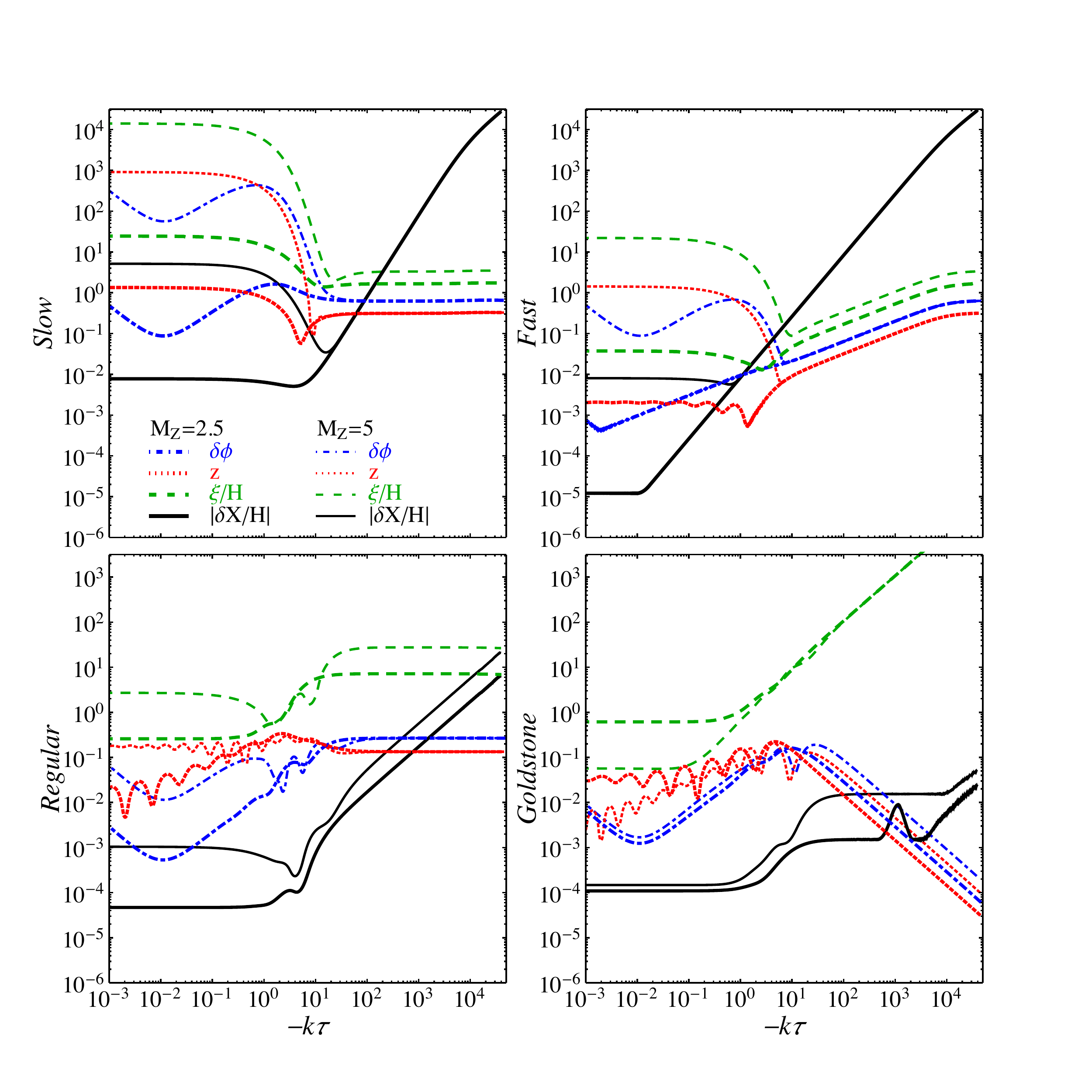}
\caption{
Evolution of scalar fluctuations in Higgsed Chromo-Natural inflation. The values of the other parameters here are chosen to be $\mu  = 8 \times 10^{-5}$, $g =  1.28\times 10^{-7}$, $\lambda \psi/f = 3920$, $\mpsi \approx 2.5$,  $H \approx 1.2 \times 10^{-9}$, $\psi \approx 0.022$. The four panels show the four independent solutions of the equations of motion, corresponding to the four independent initial conditions in eq.\ \eqref{eqn:initialconditions}. Shown are the solutions for  two different values of the Higgs vev, in thick lines $M_{\hvev} = 2.5$, and in thin lines we show $M_{\hvev} = 5$.}
\label{fig:scalarmodes}
\end{figure}

Following the treatment of \cite{vanTent:2002, Namba:2013kia}, we expand the fields into modes, 
\begin{align}\label{eqn:modeexp}
\Delta_i(\tau, {\bf k})  = \mathcal{Q}_{ij}( \tau, k)a_j({\bf k})+  \mathcal{Q}^*_{ij}(\tau, k)a^\dagger_j(-{\bf k}) , \quad \[a_i({\bf k}), a^{\dagger}_j({\bf k'})\] = \delta^{3}({\bf k} - {\bf k}')\delta_{ij},
\end{align}
where we impose the canonical commutation relation between $\Delta_i$ and its canonically conjugate momentum 
\begin{align}\label{eqn:cancom}
\[\Delta_{i}(\tau, {\bf x}), \pi_j(\tau, {\bf y}) \] = i \delta_{ij}\delta^{3}({\bf x} - {\bf y}), \quad \pi_i  \equiv \frac{\partial \mathcal{L}}{\partial (\partial_{\tau}\Delta_i^{\dagger})}.
\end{align}
Decomposing $\pi_i$ in terms of the same creation/annihilation operators as above,
\begin{align}
\pi_i(\tau, {\bf k}) = \pi_{ij}(\tau, {\bf k}) a_j({\bf k}) +\pi_{ij}^*(\tau, {\bf k})a^{\dagger}_j(-{\bf k}),  \quad \pi_{ij} = k(\mathcal{Q}'_{ij}+K_{il} \mathcal{Q}_{lj}).
\end{align}
we find that the relations (\ref{eqn:modeexp}) and (\ref{eqn:cancom}) can only be simultaneously imposed if the condition
\begin{align}\label{eqn:inicond}
\[\mathcal{Q}\pi^{\dagger} - \mathcal{Q}^* \pi^{T}\]_{ij} = i\delta_{ij}
\end{align}
is obeyed.
As pointed out by \cite{Namba:2013kia}, eq.\ (\ref{eqn:inicond}) can be imposed as an initial condition, which then holds at all times if one also imposes that the initial conditions satisfy
\begin{align}
\pi \pi^{\dagger} - \pi^* \pi^{T}  = \mathcal{Q} \mathcal{Q}^{\dagger} - \mathcal{Q}^* \mathcal{Q}^T = 0,
\end{align}
which is equivalent to imposing that the products $\pi \pi^\dagger$ and $\mathcal{Q} \mathcal{Q}^{\dagger}$ are real.

In the limit $x = -k\tau \to \infty$ the matrices $K \to \mathcal{O}(x^{-1})$ and $\Omega^2 = \mathbb{1}+\mathcal{O}(x^{-1})$ and appears that  in this limit  one can easily set the initial conditions for the mode-functions by identifying the positive frequency modes, 
\begin{align}
\mathcal{Q}_{ij} = \frac{1}{\sqrt{2k}}\delta_{ij}, \quad \mathcal{Q}'_{ij} = -i\sqrt{\frac{k}{2}}\delta_{ij}.
\end{align}
However, this turns out to be incorrect at any finite time due to the mode couplings. These become important when 
\begin{align}
x \sim \boldsymbol{\Lambda}, \text{ where } \boldsymbol{\Lambda} = \lambda \psi/f,
\end{align}
and must be retained in order to correctly identify and quantize the normal modes of the system. We now use a Wentzel-Kramers-Brillouin (WKB) method to identify and subsequently quantize the normal modes.

In the limit $x \gg  \boldsymbol{\Lambda}$, working in the slow roll limit and making use of the equations of motion for the background, we can expand the equations of motion for the fluctuations in a series in powers of $\boldsymbol{\Lambda}/x$, keeping only the terms that become important when $x \sim \boldsymbol{\Lambda}$
\begin{align}
\phihat'' + \phihat  = & \sqrt{2}\boldsymbol{\Lambda} \frac{X}{x} +\mathcal{O}\(\frac{1}{x}, \frac{\boldsymbol{\Lambda}}{x^2}\)%+ \frac{2+M_{\hvev}^2}{\mpsi} \frac{\hz}{x} 
 \\
\hz'' + \hz   = &- \sqrt{2}\mpsi \boldsymbol{\Lambda} \frac{X'}{x} +\mathcal{O}\(\frac{1}{x}, \frac{\boldsymbol{\Lambda}}{x^2}\)%+ \frac{2+M_{\hvev}^2}{\mpsi}\frac{\phihat}{x}
\\
X'' + \(1-\boldsymbol{\Lambda}^2 \frac{\mpsi^2}{x^2}\)X = &  \sqrt{2}\boldsymbol{\Lambda}\frac{\phihat}{x} + \sqrt{2}\mpsi \boldsymbol{\Lambda}\frac{ \hz'}{x}+\mathcal{O}\(\frac{1}{x},\frac{\boldsymbol{\Lambda}}{x^2}\)\\
\hat{H}'' + \hat{H} & =0+  \mathcal{O}\(\frac{1}{x}, \frac{\boldsymbol{\Lambda}}{x^2}\).
\end{align}
We adopt a WKB ansatz for the mode functions
\begin{align}
\vec{\mathcal{Q}}_{j} = \vec{a}_j \exp\[{i\int dx\, \omega(x)}\]
\end{align}
and substituting into the system of equations, neglecting terms of order $\mathcal{O}(\omega'/\omega)$ and $\mathcal{O}(\omega''/\omega)$ and we find eight solutions for the frequencies
\begin{align}\label{eq:WKBfreqs}
\omega \approx \left\{\pm 1, \pm 1, \pm \(1+\frac{3}{2}\boldsymbol{\Lambda}^2 \frac{ \mpsi^2}{x^2}\pm\frac{\boldsymbol{\Lambda}}{x}\sqrt{\frac{9}{4} \frac{\boldsymbol{\Lambda}^2}{ x^2}\mpsi^4+2(1 + \mpsi^2) }\)^{1/2}\right\}.
\end{align}
In order for the system to be stable, all of these instantaneous WKB frequencies must be real. The only frequency that is possibly imaginary is the last one for the choice of the minus sign inside the outer square root. This frequency becomes imaginary, and thus the mode becomes unstable when
\begin{align}
1+\frac{3}{2}\boldsymbol{\Lambda}^2 \frac{ \mpsi^2}{x^2} < \frac{ \boldsymbol{\Lambda} }{x}\sqrt{\frac{9}{4} \frac{\boldsymbol{\Lambda}^2}{x^2} \mpsi^4+2 (1+ \mpsi^2)},
\end{align}
which occurs when
\begin{align}
x^2 <  \boldsymbol{\Lambda}^2 \left(2-\mpsi^2\right).
\end{align}
Thus there is an instability in the system for parameters such that  $m_{\psi}^2 = g^2\psi^2/H^2 < 2$, as was found for the original model \cite{Dimastrogiovanni:2012ew, Adshead:2013nka, Adshead:2013qp}. The additional mass for the gauge fields does not stabilize the low mass regime, at least in the limit $M_{\hvev} \ll  \boldsymbol{\Lambda}$.

Expanding the eigenfrequencies in eq.\ \eqref{eq:WKBfreqs} in the limit $x \to \infty$, we have
\begin{align}
\omega \sim \left\{\pm 1, \pm 1, 1\pm  \boldsymbol{\Lambda} \frac{\sqrt{1+\mpsi^2}}{\sqrt{2}x}, -1\pm  \boldsymbol{\Lambda} \frac{\sqrt{1+\mpsi^2}}{\sqrt{2}x}\right\}
\end{align}
and the corresponding mode solutions are, up to an irrelevant phase
\begin{align}\nn
\vec{\mathcal{Q}}_{j} = & c_{1j}\vec{a}_{1}e^{-i x}+c_{2j} \vec{a}_{2}e^{ix} +c_{3j} \vec{a}_{3}e^{-i x}+c_{4j} \vec{a}_{4}e^{ix}+c_{5j} \vec{a}_{5}e^{-i x +  i\boldsymbol{\Lambda} \frac{\sqrt{1+\mpsi^2}}{\sqrt{2}}\ln x}\\ &+c_{6j} \vec{a}_{6}e^{i x -  i  \boldsymbol{\Lambda}\frac{\sqrt{1+\mpsi^2}}{\sqrt{2}}\ln x}  +c_{7j}\vec{a}_{7} e^{-i x -   i\boldsymbol{\Lambda}\frac{\sqrt{1+\mpsi^2}}{\sqrt{2}}\ln x}  +c_{8j} \vec{a}_{8}e^{i x+  i \boldsymbol{\Lambda} \frac{\sqrt{1+\mpsi^2}}{\sqrt{2}}\ln x}
\end{align}
where the $c_{ij}$ are constants and the $\vec{a}_{i}$ are the vectors
\begin{align}
\vec{a}_{1} = \vec{a}_2 = &
\left[
\begin{array}{ccc}
0    \\
0 \\
0  \\
1   
\end{array}
\right],
\quad \quad\quad\quad 
\vec{a}_{3} = \vec{a}_4^* = 
\left[
\begin{array}{ccc}
\frac{i\sqrt{2} \mpsi^2 }{2+M_{\hvev}^2} \boldsymbol{\Lambda}  \\
\frac{\sqrt{2} \mpsi }{2+M_{\hvev}^2} \boldsymbol{\Lambda} \\
1  \\
0   
\end{array}
\right],\\
\vec{a}_{5} = \vec{a}_6^* = &
\left[
\begin{array}{ccc}
\frac{1}{\sqrt{1+\mpsi^2}} \\
\frac{i\mpsi}{\sqrt{1+\mpsi^2}}\\
1  \\
0   
\end{array}
\right], \quad
\vec{a}_7 = \vec{a}_{8} ^*= 
\left[
\begin{array}{ccc}
-\frac{1}{\sqrt{1+\mpsi^2}} \\
-\frac{i\mpsi}{\sqrt{1+\mpsi^2}}\\
1  \\
0   
\end{array}
\right].
\end{align}
Note that in all cases, in finding both the frequencies and vectors we have expanded and dropped terms that are subleading both in powers of $\boldsymbol{\Lambda}$ and $x = -k\tau$.

Now, demanding the solutions approach the positive frequency solutions as $x = -k\tau \to \infty$ means we can set $c_{1j} = c_{3j} = c_{5j} = c_{7j} = 0$. The remaining constants now need to be set be imposing the quantization conditions above. Working in the limit $x\to\infty$, it is then straightforward to see that a solution that satisfies the initial conditions is
\begin{align}\label{eqn:initialconditions}
  \text{ Goldstone mode:} \quad c_{21} =  & c_{22} = c_{33} = 0, \quad c_{24} = \frac{1}{\sqrt{2k}}\\
 \text{ Regular mode:} \quad c_{42} = &  c_{43} = c_{44} = 0,\quad c_{41} = \frac{1}{\sqrt{2k}}\frac{1}{\sqrt{1+2 \boldsymbol{\Lambda}^2\frac{\mpsi^2(1+\mpsi^2)}{(2+M_{\hvev}^2)^2}}} \\
 \text{ Slow mode:} \quad c_{61} =  & c_{62} = c_{64} = 0, \quad c_{63} = \frac{1}{2\sqrt{k}}\\
 \text{ Fast mode:} \quad c_{81} = &  c_{83} = c_{84} = 0, \quad c_{82} = \frac{1}{2\sqrt{k}}.
\end{align}

We show the solutions to all four modes in figure \ref{fig:scalarmodes}. The gauge field fluctuation  $z$ becomes constant at late times, while $\delta\phi$ grows. However, note that the physical gauge field fluctuations are proportional to $\delta\phi/a$, which becomes constant at late times. In principle, to numerically solve the system of equations and obtain all solutions, one needs to solve the system of equations four times starting the system in each of the four normal modes with the other amplitudes set to zero. However, in practice we are interested in the curvature fluctuation which, as we demonstrate below, to a good approximation arises solely from the axion fluctuation.  Note that in all but the ``slow-mode'', which corresponds to the magnetic drift mode with $c_{63} \neq 0$, the axion fluctuation has a negligible final amplitude. Thus to a very good approximation we need only simulate this mode when computing the curvature spectrum. 

Notice that the effect of the Higgs VEV and accompanying Goldstone fluctuations boosts the final amplitude of the fluctuations. These dynamics are what allows the model to become consistent with the data -- the scalar curvature fluctuations are boosted, {more than the tensor fluctuations,} thus lowering the tensor-to-scalar ratio. In generating figure \ref{fig:scalarmodes}, we have smoothed the curves for the fast, regular and Goldstone modes to eliminate contamination from errors in the initial conditions. This smoothing is not required for the slow mode and the initial condition above is an excellent approximation.

\section{Tensors}

We now turn to the spin-2 modes. While the addition of the Higgs sector introduces new scalar and vector degrees of freedom via the Goldstone modes, in this limit there are no new spin-2 degrees of freedom in the theory. The only difference in the case at hand from Chromo-Natural Inflation is that the Higgs generates a new mass term for the spin-2 parts of the gauge field and the graviton, and changes the details of the interactions between them. We thus expect the analysis of these modes to mirror that of Chromo-Natural Inflation presented in refs.\ \cite{Adshead:2013nka, Adshead:2013qp}. The action for the canonical variables
\begin{align}
\hg^\pm = \frac{a \gamma^\pm}{\sqrt{2}}, \text{ and } \hy^\pm = \sqrt{2}t^\pm,
\end{align}
is given by
\begin{align}\nn
\mathcal{S} = \frac{1}{2}\int \frac{d^3 k}{(2\pi)^3} d\tau \Bigg[&
\partial_{\tau}\hg_k^{\pm}\partial_{\tau}\bar{\hg}_k^{\pm}-\(k^2-\frac{1}{a}\frac{\partial^2 a}{\partial\tau^2}-2\dot{\phi}^2 + 2 g^{2}\frac{\phi^4}{a^2}+2g^2a^2\hvev^2\psi^2\)\hg_k^{\pm}\bar{\hg}_k^{\pm}\\ \nn & 
+\partial_{\tau}\hy_k^{\pm}\partial_{\tau}\bar{\hy}_k^{\pm}-\(k^2 +  g\phi\frac{\lambda}{f}\partial_{\tau}\axion+2g^ 2 a^2  \Phi^2  \)\hy_k^{\pm}\bar{\hy}_k^{\pm}  \pm k\(\frac{\lambda}{f}\partial_{\tau}\axion + 2g\phi\) \hy_k^{\pm}\bar{\hy}_k^{\pm}
\\\nn &  -  2\dot{\phi}( \partial_{\tau} \hy^{\pm}_k\bar{\hg}_k^{\pm}+\partial_{\tau} \bar{\hy}_k^{\pm}{\hg}_k^{\pm}) \mp 2k g\frac{\phi^2}{a} (\hy_k^{\pm}\bar{\hg}_k^{\pm}+\bar{\hy}_k^{\pm}{\hg}_k^{\pm}) \\ & +2g^2\frac{\phi^3}{a} (\hy_k^{\pm}\bar{\hg}_k^{\pm}+\bar{\hy}_k^{\pm}{\hg}_k^{\pm}) +2g^2a^2\hvev^2\psi (\bar{\hy}^{\pm} \hg^{\pm} + \hy^{\pm} \bar{\hg}^{\pm})\Bigg].
\end{align}
Note that these modes are not subject to either the constraints from the Einstein equations or the Gauss law constraints at linear order in perturbation theory. Furthermore, they are invariant under coordinate and SU(2) gauge transformations. We work in the slow roll limit, and introducing $x = -k\tau$, this action becomes
\begin{align}\nn
\mathcal{S} = \frac{1}{2}\int \frac{d^3 k}{(2\pi)^3} \frac{dx}{-k} \Bigg[&
\partial_x\hg_k^{\pm}\partial_x\bar{\hg}_k^{\pm}-\(1-\frac{2}{x^2}+\frac{2\psi^2}{x^2}\(\mpsi^2 -1+m_{\hvev}^2\)\)\hg_k^{\pm}\bar{\hg}_k^{\pm}\\ \nn & 
+\partial_{x}\hy_k^{\pm}\partial_{x}\bar{\hy}_k^{\pm}-\(1 +  \frac{\mpsi}{x^2}\frac{\lambda}{f}\frac{\dot\axion}{H}+\frac{M_{\hvev}^2}{x^2}  \)\hy_k^{\pm}\bar{\hy}_k^{\pm}  \pm \frac{1}{x}\(\frac{\lambda}{f}\frac{\dot\axion}{H} + 2\mpsi\) \hy_k^{\pm}\bar{\hy}_k^{\pm}
\\\nn &  + \frac{ 2\psi}{x}( \partial_{x} \hy^{\pm}_k\bar{\hg}_k^{\pm}+\partial_{x} \bar{\hy}_k^{\pm}{\hg}_k^{\pm}) \mp 2 \psi\frac{\mpsi}{x} (\hy_k^{\pm}\bar{\hg}_k^{\pm}+\bar{\hy}_k^{\pm}{\hg}_k^{\pm}) \\ & +2\psi \frac{\(\mpsi^2+M_{\hvev}^2\)}{x^2} (\hy_k^{\pm}\bar{\hg}_k^{\pm}+\bar{\hy}_k^{\pm}{\hg}_k^{\pm}) \Bigg],
\end{align}
Varying the action, we find the equations of motion for the fields,
\begin{align}\label{eqn:grav}
\hg_k^{\pm}{}''+\(1-\frac{2}{x^2}+\frac{2\psi^2}{x^2}\(\mpsi^2 -1+M_{\hvev}^2\)\)\hg_k^{\pm} =  \frac{ 2\psi}{x}\partial_{x} \hy^{\pm}_k  \mp 2 \psi\frac{\mpsi}{x} \hy_k^{\pm}+2\psi \frac{\(\mpsi^2+M_{\hvev}^2\)}{x^2} \hy_k^{\pm}
\end{align}
and
\begin{align}\nn\label{eqn:gauge2}
\hy_k^{\pm}{}''+\(1 +  \frac{\mpsi}{x^2}\frac{\lambda}{f}\frac{\dot\axion}{H}+\frac{M_{\hvev}^2}{x^2}\)\hy_k^{\pm}  \mp \frac{1}{x}\(\frac{\lambda}{f}\frac{\dot\axion}{H} + 2\mpsi\) \hy_k^{\pm} = & -  2\psi \partial_{x} \frac{\bar{\hg}_k^{\pm}}{x} + 2 \psi    \left ( {1\over x^2}\mp \frac{\mpsi}{x} \right ) {\hg}_k^{\pm}    \\ &+2\psi \frac{\(\mpsi^2+M_{\hvev}^2\)}{x^2} {\hg}_k^{\pm}
\end{align}
Notice that the spin-2 mode of the gauge field becomes temporarily unstable due to the fact that its instantaneous WKB frequency  becomes temporarily negative. Making use of the background equations of motion (\ref{slowrollpsi})  and (\ref{slowrollX}), and taking $\mpsi , M_{\hvev} \gg 1$ we find that the instantaneous WKB frequency is negative during the period,
\begin{align}\label{eqn:tensorinstability}
\frac{4 \mpsi^2+M_{\hvev}^2+\sqrt{8 \mpsi^4+M_{\hvev}^4}}{2\mpsi} \gtrsim x \gtrsim  \frac{4 \mpsi^2+M_{\hvev}^2-\sqrt{8 \mpsi^4+M_{\hvev}^4}}{2 \mpsi}, 
\end{align}
during which period the amplitude of the gauge field will increase exponentially. Note that, although we have given the gauge fields an additional mass term, this then causes the background axion to roll faster which means the instability is still present. However, while this ruled out the previous models of Chromo-Natural Inflation and Gauge-flation, the addition of the Goldstone modes alters the scalar dynamics in such a way as to allow this model to satisfy current observational constraints.

\begin{figure}[t!]
\centering
\includegraphics[width=\textwidth]{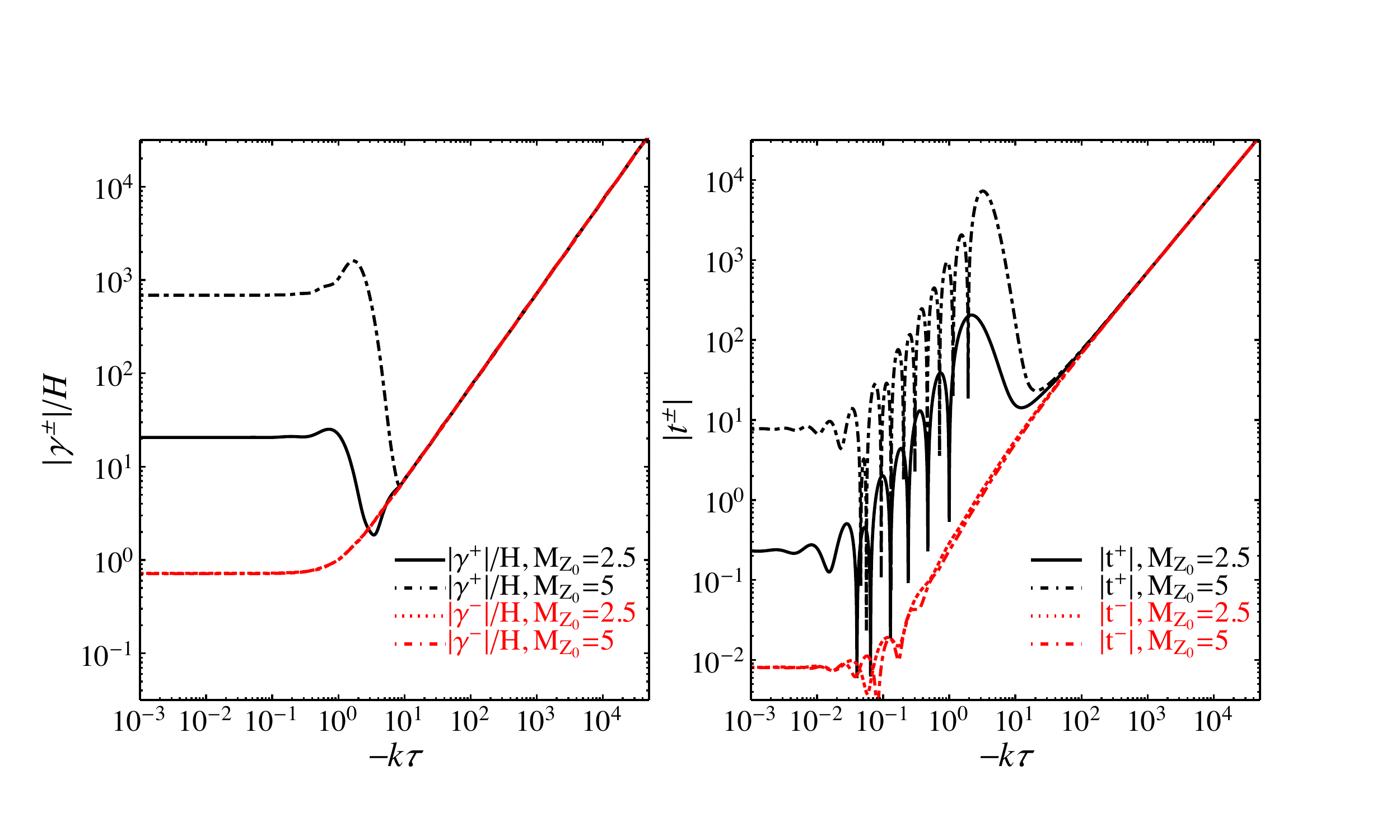}
\caption{
Evolution of tensor fluctuations in Higgsed Chromo-Natural inflation. The addition of the Higgs leads to a temporary exponential instability in the tensor modes that begins near $k\tau \sim M_{Z_0}$. The values of the other parameters here are chosen to be $\mu =  8 \times 10^{-5}$, $g =  
 1.28  \times 10^{-7}$, $\lambda \psi/f = 3920$, $\mpsi \approx 2.5$,  $H \approx 1.2 \times 10^{-9}$, $\psi \approx 0.022$.
 }
\label{fig:tensormodes}
\end{figure}

In figure \ref{fig:tensormodes} we plot the evolution of the tensor modes for this model. Note that one helicity of the gauge tensor is strongly amplified, which in turn strongly amplifies one of the gravitational wave helicities. We also note that the gauge tensor appears to freeze out on large scales. However, this does not lead to any contribution to the stress tensor at late times, as it contributes as $t^{\pm}/a$, and thus its contributions decay at late times.

\subsection{Approximate solutions}

Analogously to Chromo-Natural inflation, the above set of equations, \eqref{eqn:grav} and \eqref{eqn:gauge2}, admits an excellent analytic approximation. Note first that the coupling between these equations proportional to $\psi \ll 1$, while the dominant part of the gauge field equation of motion is its mass term. For the left handed $\hy^{+}$ modes, this mass becomes negative leading to the exponential enhancement of its amplitude;  the right-handed modes remain stable. This suggests that, to a good approximation, we may simply ignore the right-handed gauge field modes and treat the right-handed gravitational wave modes as unperturbed. For the left-handed modes we can solve the free equation of motion for the gauge field modes and use these solutions as sources for the left-handed gravitational wave equation of motion. Furthermore, we may neglect the mass terms for the graviton,\footnote{Note that this small mass term will lead to evolution of the gravitational wave amplitudes on super-horizon scales, and may lead to interesting effects in the tensor squeezed limits \cite{Bordin:2016ruc}} treating its equation of motion as that of a massless scalar field in de Sitter space. In this approximation, the equation of motion for the gauge field is well approximated by
\begin{align}
\hy_k^{+}{}''+\(1 +  \frac{m^2}{x^2} - \frac{m_t}{x}\)\hy_k^{+}  = 0,
\end{align}
where we have defined
\begin{align}
m = & 2(1+\mpsi^2 +M_{\hvev}^2) =   \frac{1}{4}-\beta^2\\
m_t = & \frac{1}{\mpsi}\(2+4\mpsi +M_{\hvev}^2\)  = -2i \alpha,
\end{align}
introducing $\alpha$ and $\beta$ for convenience. The modes can be quantized in an analogous fashion to that presented above. The analysis of this case is identical to that of Chromo-Natural inflation \cite{Adshead:2013nka, Adshead:2013qp}, and we refer the reader to those works for details and merely state the results here.

At late times, the solution for the left-handed gravitational wave is well approximated by
\begin{align}\label{eqn:gammaplus}
\gamma^{+}(x) = &   \frac{H x}{\sqrt{k^3}} u_1 (x)+2\sqrt{2}\frac{H}{k} B_k \psi  \(I_1+\mpsi I_2 -(\mpsi^2+M_{\hvev}^2) I_3\),
\end{align}
where $u_1(x)$ is the free solution of the canonically normalized gravitational wave equation,
\begin{align}
u_{1}(x) = & \(1+\frac{i}{x}\)e^{i x}
\end{align}
and
\begin{align}\nn
I_1 = & \frac{\left(m^2-2 i m m_t+2 m-2 m_t^2\right) \sec \left( \pi \beta \right) \sinh \left(-i\pi\alpha\right) \Gamma \left(\alpha\right)}{2 m (m+2)} \\ \nn&-\frac{\pi ^2 \left(m^2+2 i m m_t+2 m-2 m_t^2\right) \sec \left( \pi \beta \right) \text{csch}\left(-i\pi\alpha\right)}{2 m (m+2) \Gamma \left(\alpha+1\right) \Gamma \left(-\alpha-\beta+\frac{1}{2}\right) \Gamma \left(-\alpha+\beta+\frac{1}{2}\right)}~,\\
\nn
I_2 = & \frac{\pi  \sec \left(\pi \beta\right) \Gamma \left(-\alpha\right)}{2 \Gamma \left(-\alpha-\beta+\frac{1}{2}\right)
   \Gamma \left(-\alpha+\beta+\frac{1}{2}\right)}
   -\frac{\pi  \sec \left(\pi \beta\right) \Gamma \left(1-\alpha\right)}{m \Gamma \left(-\alpha-\beta+\frac{1}{2}\right) \Gamma \left(-\alpha+\beta+\frac{1}{2}\right)} \\ \nn&
    +\frac{\pi  m \sec \left( \pi  \beta\right)-i \pi  m_t \sec \left( \pi  \beta\right)}{2 m \Gamma \left(1-\alpha\right)} ~,
\nn\\
I_3 = &\frac{\pi ^2 (m+i m_t) \text{sec} \left(\pi 
   \beta\right) \text{csch}\left(-i\pi\alpha \right)}{m (m+2) \Gamma \left(\alpha \right) \Gamma \left(-\alpha-\beta+\frac{1}{2}\right) \Gamma \left(-\alpha+\beta+\frac{1}{2}\right)} +  \frac{\pi  (m_t+i m) \text{sec} \left( \pi\beta\right)}{m (m+2) \Gamma \left(-\alpha\right)}~.
\end{align}
The total gravitational wave power spectrum at late times, $k\tau_* \to 0$, is given by
\begin{align}
\Delta^{2}_{\gamma}(k) =2 \Delta^{2}_{\gamma^+}(k)+2\Delta^2_{\gamma^-}(k) 
\end{align}
where the spectra of left and right-handed modes are defined by
\begin{align}
\langle \gamma^{\pm}_{\bf k}(\tau_*)\gamma^{\pm}_{\bf k'}(\tau_*) \rangle= (2\pi)^3\delta^{3}({\bf k}+{\bf k'})\frac{2\pi^2}{k^{3}}\Delta^{2}_{\gamma^{\pm}}(k) .
\end{align}
Now, the right handed modes $\gamma^{-}$ are, to a very good approximation, unaffected by their interactions with the spin-2 fluctuations of the gauge fields.  Their spectrum is given by the usual result,
\begin{align}\label{eqn:rhgravspec}
\Delta^{2}_{\gamma^-}(k) = \frac{ H^2}{2\pi^2}.
\end{align}
For the left handed modes, $\gamma^+$, the vacuum fluctuations are uncorrelated with the contribution due to their interaction with the gauge field fluctuations, and thus to a good approximation 
\begin{align}\label{eqn:lefthandedgw}
\Delta^{2}_{\gamma^+}(k) = \frac{ H^2}{2\pi^2}+4k \frac{H^2}{\pi^2}\psi^2  |B_k|^2  |I_1+\mpsi I_2 -(\mpsi^2+M_{\hvev}^2) I_3|^2.
\end{align}

\begin{figure}
\centerline{\includegraphics[width=.75\textwidth]{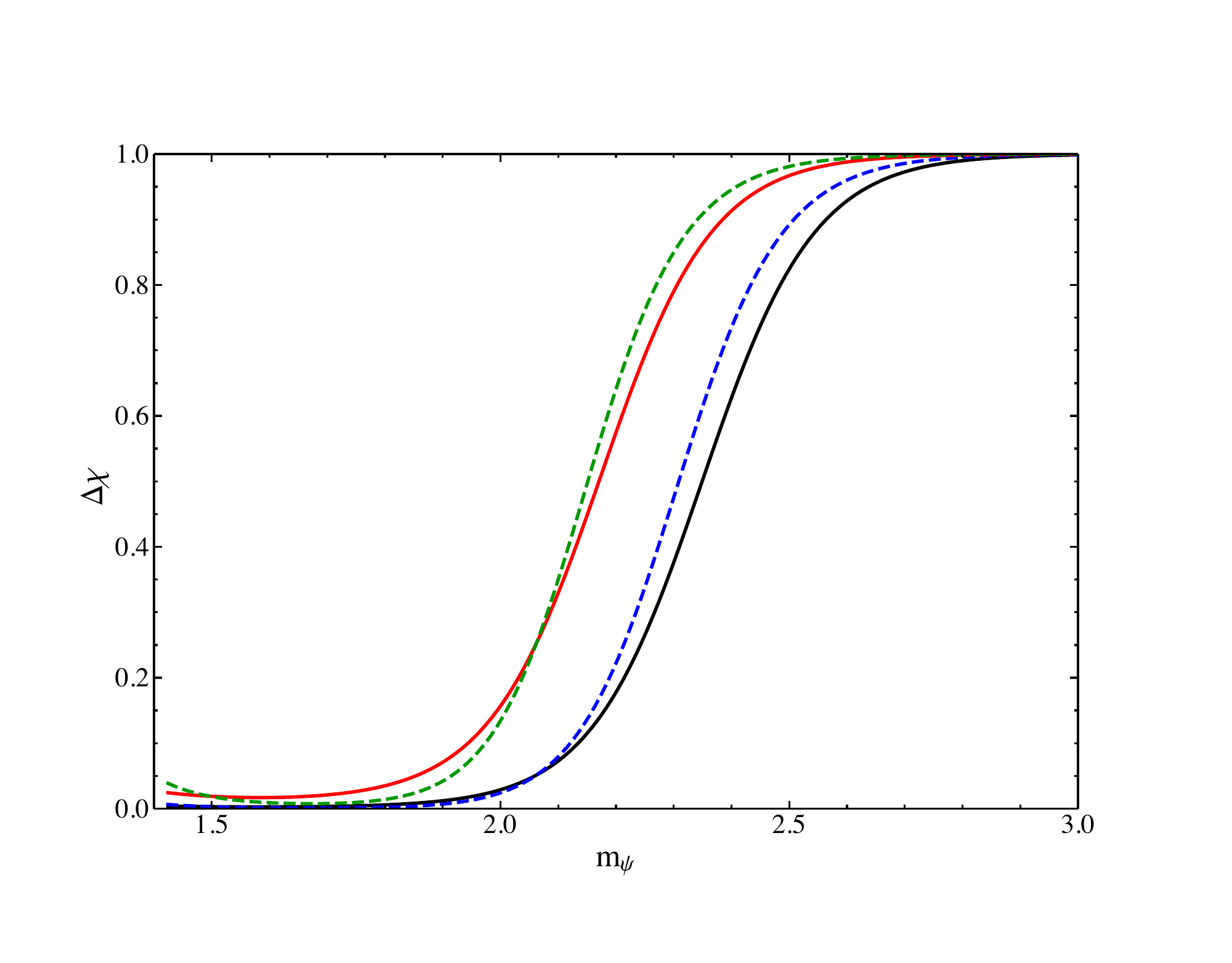}}
\caption{
The chirality parameter $\Delta \chi$ as a function of $m_\psi$ for two different values of the gauge field VEV $\psi=0.01$ (blue and black) and $\psi=0.025$ (red and green). The Higgs mass is set to $M_{Z_0}=0$ (solid curves) and $M_{Z_0}=2$ (dashed curves).
}
\label{fig:DeltaChi}
\end{figure}

We can also compute the chirality parameter
\begin{align}
\Delta\chi = \frac{\Delta^{2}_{\gamma^+} - \Delta^{2}_{\gamma^-}}{\Delta^{2}_{\gamma^+} + \Delta^{2}_{\gamma^-}}.
\end{align}
This quantity is plotted in figure \ref{fig:DeltaChi}. Note that the resulting gravitational wave spectrum very quickly becomes completely polarized as $m_\psi$ is increased.

%%%%%%%%%%%%%%%%%%%%%%%%%%%%%%%%%%%%
%%%%%%%%%%%%%%%%%%%%%%%%%%%%%%%%%%%%
%%
\section{Curvature perturbations and primordial spectra}\label{sec:curvature}
%%
%%%%%%%%%%%%%%%%%%%%%%%%%%%%%%%%%%%%
%%%%%%%%%%%%%%%%%%%%%%%%%%%%%%%%%%%%

In this section we consider the effect of field fluctuations on the late time universe. The presence of multiple degrees of freedom means that the curvature fluctuations on large scales may evolve due to the presence of entropy perturbations. We begin by evaluating the curvature fluctuation on superhorizon scales. We then calculate the entropy perturbation, demonstrating that it is slow-roll suppressed compared to the adiabatic curvature fluctuation.

\subsection{Curvature perturbations}

The comoving curvature perturbation is given by the gauge invariant quantity
\begin{align}
\curv = \frac{A}{2} + H\delta u,
\end{align}  
where $\delta u$ is the perturbation to the scalar velocity potential, and $A$ is defined via the general perturbed spatial metric
\begin{align}
\tilde{h}_{ij} = a^2\[(1+A)\delta_{ij} + \partial_i \partial_j B+ \partial_i C_j + \partial_j C_i + \gamma_{ij}\].
\end{align}
In spatially flat gauge, $A = B = 0$ by coordinate choice and thus, it remains to find the perturbation to the velocity potential. This is found from the perturbation to the momentum flux
\begin{align}
\delta T_{0i} = & \bar{p}\, \tilde{h}_{i0} - a (\bar\rho+\bar p)(\partial_i \delta u + \delta u_{V}^i).
\end{align}
In this expression, 
$\delta u$ and $\delta u^i_V$ are the scalar and vector perturbations to the velocity potential respectively, while $\tilde{h}_{0i} = a^{2}\delta_{ij}N^j$ is the perturbation to the space-time components of the metric $g_{\mu\nu}$ and $\bar\rho$ and $\bar p$ are the background energy density and pressure respectively. 

The stress tensor for the theory defined at eqs.\ (\ref{eqn:CNIaction}) and (\ref{eqn:Higgsaction}) is given by
\begin{align}\nn
T_{\mu\nu} = & 2\tr\[F_{\mu\alpha}F_{\nu\beta}\]g^{\alpha\beta}-\frac{g_{\mu\nu}}{2}\tr\[F_{\alpha\beta}F^{\alpha\beta}\]+ \partial_{\mu}\axion\partial_{\nu}\axion -g_{\mu\nu}\left[\frac{1}{2}g^{\rho\sigma}\partial_{\rho}\axion\partial_{\sigma}\axion+V(\axion)\right]\\
&    +2g^2\hvev^2\tr\[ (A_{\mu} - \frac{i}{g}U^{-1}\partial_\mu U)(A_{\nu} - \frac{i}{g}U^{-1}\partial_\nu U)\] - g_{\mu\nu}g^2\hvev^2\tr\[ (A_{\mu} - \frac{i}{g}U^{-1}\partial_\mu U)^2\].
\end{align}
We can calculate the momentum flux from this expression; to linear order in fluctuations, it is
\begin{align}
T_{0i} 
= & (\bar{p}_{\axion}+\bar{p}_{YM}+\bar{p}_{Z})a^{2}\delta_{ij}N^j+a\dot\axion\partial_{i}\daxion
 + \frac{\dot{\phi}}{a}(2\partial_{[k}\da^k_{i]} + g \phi \epsilon_{aik}   \da^a_k)\\\nn
 &-g\frac{\phi^2}{a^2}(  \epsilon^{a}_{ki}\partial_\tau \da^a_k -  \epsilon^{a}_{ik}\partial_{k}\da^a_0 -2 g \phi \da^i_{0})+g^2\hvev^2 a\psi \delta_{ai} \da^a_0 + g \hvev^2 a\psi \partial_0 \dHiggs^i  \\ &  -\(2g^2\psi^4+\frac{1}{2}g^2\hvev^2\psi^2\)a^2\delta_{ij}N^j .
\end{align}
Inserting our field decomposition and ignoring the vector degrees of freedom we find
\begin{align}\nn
T_{0i} 
\approx & (\bar{p}_{\axion}+\bar{p}_{YM}+\bar{p}_{Z})a^{2}\delta_{ij}N^j+a\dot\axion\partial_{i}\daxion
  -\(H\psi \partial_{i}(2 z+4\delta\phi) - g a \psi^3\frac{\lambda}{f}\partial_{i}\daxion\)\\ &%+ g \hvev^2 a\psi \partial_0 \dHiggs^i 
  -\(2g^2\psi^4+g^2\hvev^2\psi^2\)a^{2}\delta_{ij}N^j \,,
  \end{align}
where the `$\approx$' indicates that we have dropped terms that decay at late times and worked in the slow-roll approximation. We have also made use of the non-Abelian gauge condition and imposed the Gauss's law constraint, eq.\ (\ref{eqn:scalargauss}), in the long wavelength limit ($k \to0$). In this expression, $\bar{p}_{\axion}$, $\bar{p}_{\rm YM}$ and $\bar{p}_Z$ are the background isotropic pressures due to the axion, gauge fields and Higgs respectively.

Now, note that in the limit $k\to 0$ where $\lambda \gg 1$, to a very good approximation the curvature perturbation is given by
\begin{align}
\curv \approx \frac{H}{\bar{\rho}+\bar{p}}g \psi^3 \frac{\lambda}{f}\daxion = \frac{\mpsi \psi^2}{2  \epsilon_H} \frac{\lambda}{f}\daxion,
\end{align}
where $\bar{\rho}$ and $\bar{p}$ are the total background energy density and pressure respectively. In the second equality we have made use of the relation
\begin{align}
\epsilon_H = \frac{\bar{\rho}+\bar{p}}{2H^2}.
\end{align}
Thus the form of the curvature fluctuation is identical to that of Chromo-Natural inflation, and admits the familiar interpretation from single clock inflation that the inflaton is simply acting as a clock. The curvature perturbation then arises as fluctuations of the time on this clock from place to place
\begin{align}
\curv \approx \frac{\daxion}{\Delta\axion}
\end{align}
where $\Delta\axion = \dot\axion/H$.

\subsection{Entropy perturbations and isocurvature}

As noted above, the additional degrees of freedom in the theory leave open the question of how these fluctuations affect the curvature perturbation. To address this, we compute the entropy perturbation, a gauge invariant quantity defined (see e.g. \cite{Wands:2000dp})
\begin{align}
\mathcal{S} = H\(\frac{\delta p }{\dot p} - \frac{\delta\rho}{\dot\rho}\) = \frac{H}{\dot\rho}\(c_{s}^{-2}\delta p  - \delta\rho\) .
\end{align}
The quantity in parenthesis is proportional to the total non-adiabatic pressure perturbation.  In Higgsed Chromo-Natural Inflation, the background energy and pressures are given by
\begin{align}
\rho = & \frac{1}{2}\dot\axion^2 + V(\axion) + \frac{3}{2}\(\frac{\dot\phi^2}{a^2} + g^2 \frac{\phi^4}{a^4}\) + \frac{3}{2} g^2 \psi^2 \hvev^2\\
p = & \frac{1}{2}\dot\axion^2 - V(\axion) + \frac{1}{2}\(\frac{\dot\phi^2}{a^2} + g^2 \frac{\phi^4}{a^4}\) - \frac{1}{2}g^2 \psi^2 \hvev^2.
\end{align}
Differentiating these expressions, to a good approximation we then find
\begin{align}
\frac{\dot{\rho}}{H} \approx& 3H^2\(2+2\mpsi^2 + \mHiggs^2\)\epsilon_\psi\(1+ \eta_{\psi} - \frac{2}{2+2\mpsi^2 + \mHiggs^2} \epsilon_H\)\\\
\frac{\dot{p}}{H} \approx &-3H^2\(2+2\mpsi^2 +\mHiggs^2 \)\epsilon_\psi\(1-\frac{1}{3}\frac{(2+2\mpsi^2 - \mHiggs^2)\eta_{\psi} - 2\epsilon_H }{2+2\mpsi^2 + \mHiggs^2} \)
\end{align}
where we have used the background equations of motion, eqs.\ (\ref{slowrollpsi}) and (\ref{slowrollX}) and dropped the terms involving $\ddot{\axion}$ and $\ddot{\psi}$. The adiabatic sound speed is
\begin{align}
c_{s}^2 = \frac{\dot{p}}{\dot\rho}  \approx  -1+\frac{2}{3}\frac{(2+4\mpsi^2 +\mHiggs^2)\eta_\psi - 4 \epsilon_H}{2+2\mpsi^2 + \mHiggs^2} .
\end{align}
The energy density and pressure at linear order in field fluctuations are found by evaluating
\begin{align}
\rho  = - T^{0}{}_{0}, \quad p  = \frac{\delta^{j}{}_{i}}{3}T^{i}{}_{j},
\end{align}
where $T^{\mu}{}_{\nu}$ is the mixed energy-momentum tensor. In spatially flat gauge, the density and pressure perturbations are
\begin{align}\nn
\delta\rho = & V' \daxion + \(\dot\axion\dot\daxion-\delta N a^2\dot\axion^2\)+\frac{1}{a^2}\hvev^2\(g\phi \partial_i \dHiggs^i+ g^2 \phi \da^i_{i}\) -3 \delta N \dot\phi^2  \\& 
+\frac{\dot\phi}{a^3}(\partial_{\tau}\da^i_{i} - \partial_j \da^j_0) +\frac{g\phi^2}{a^{4}}( \epsilon^a_{mi} \partial_{[m}\da^{a}_{i]}+2g\phi \da^i_i )\\\nn
\delta p  = & -V' \daxion + \(\dot\axion\dot\daxion-\delta N a^2\dot\axion^2\)-\frac{1}{3a^2}\hvev^2\(g\phi \partial_i \dHiggs^i+ g^2 \phi \da^i_{i}\) -\delta N \dot\phi^2 \\& 
+\frac{\dot\phi}{3a^3}(\partial_{\tau}\da^i_{i} - \partial_j \da^j_0) +\frac{g\phi^2}{3a^{4}}( \epsilon^a_{mi} \partial_{[m}\da^{a}_{i]}+2g\phi \da^i_i ).
\end{align}
where $\delta N$ is the first order perturbation to the lapse function in eq.\ (\ref{eqn:adm}).  Working in the long wavelength limit, dropping the metric terms (which are expected to be small, in addition to having small coefficients), we find the entropy perturbation
\begin{align}
\mathcal{S}  \approx   - \frac{\psi H \tau }{3 \epsilon_H }\( 2 \frac{ \partial \delta\phi }{\partial \ln \tau} +\(4 \mpsi^2 +\mHiggs^2\)\delta\phi \)-\frac{2}{3}\frac{\(4+8\mpsi^2 +2\mHiggs^2\)\eta_\psi -4\epsilon_H }{2+2\mpsi^2 + \mHiggs^2} \frac{ g\psi^3}{2 H \epsilon_H}\frac{\lambda}{f}\daxion.
\end{align}
Note that, ignoring the fluctuations from the gauge fields for a moment, the ratio of the entropy perturbation to the adiabatic curvature perturbation due to the axion fluctuations in the long wavelength limit is proportional to the small quantities $\epsilon_H, \eta_{\psi} \ll 1$, and thus the ratio of their power goes as the square of this quantity. Further, note that the contribution to the entropy perturbation due to the gauge field fluctuations is proportional to $\psi$. Thus, this contribution is additionally suppressed since the final gauge field fluctuations are small compared to those of the axion. 

The presence of a non-zero entropy perturbation $\mathcal{S}$ will cause the adiabatic mode to evolve on large scales, 
%The significance of this quantity is that it leads to evolution of the curvature fluctuation on super-horizon scales
\begin{align}
\dot{\curv} = -3H\frac{\dot{p}}{\dot\rho}\mathcal{S}.
\end{align}
However, this effect will be comparable to the effect of the gravitational interactions which we have neglected in this work. In what follows, we will evaluate the curvature perturbation near horizon crossing, and postpone analysis of the super-horizon evolution to future work.

\subsection{Density fluctuation}

As a check we may also evaluate the curvature fluctuation on uniform density hypersurfaces
\begin{align}
\zeta = \frac{A}{2} - \frac{H\delta\rho}{\dot{\rho}}
\end{align}
On superhorizon scales, where we can neglect gradients, $\zeta$ and $\curv$ should agree. As a non-trivial check we can test this.
%\begin{align}
%\curv = \zeta
%\end{align}
From above, the perturbation to the energy density is dominated by the axion
\begin{align}
\delta\rho \approx V' \daxion \approx -3 g H \psi^3 \frac{\lambda}{f} \daxion
\end{align}
so that, after using the continuity equation for the background fluids, we have in spatially flat gauge in the long wavelength limit
\begin{align}
\zeta = - \frac{H\delta\rho}{\dot{\rho}}  \approx & \frac{3 g H^2 \psi^3}{\dot\rho}  \frac{\lambda}{f}\daxion = \frac{g\psi^3}{2 H \epsilon_H} \frac{\lambda}{f}\daxion
\end{align}
and thus we confirm that
%\begin{align}
$\curv = \zeta$
%\end{align}
 in the long wavelength limit.

%%%%%%%%%%%%%%%%%%%%%%%%%%%%%%%%%%%%
%%%%%%%%%%%%%%%%%%%%%%%%%%%%%%%%%%%%
%%
\section{Parameter dependence and observational constraints}\label{sec:obs}
%%
%%%%%%%%%%%%%%%%%%%%%%%%%%%%%%%%%%%%
%%%%%%%%%%%%%%%%%%%%%%%%%%%%%%%%%%%%

The system of equations for the fluctuations in Higgsed Chromo-Natural Inflation is extremely involved, and does not readily admit analytic solutions. Furthermore, our study of the tensor sector of this model (c.f.\ eq.\ (\ref{eqn:tensorinstability})) showed that the Higgs mechanism does not entirely quell the strong gravitational wave production that prevented the earlier version of this model from agreeing with data. In order to determine if the model is viable, we perform a numerical scan of the parameter space to determine the scalar and tensor power spectra, as well as their tilts.

\begin{figure}
\includegraphics[width=\textwidth]{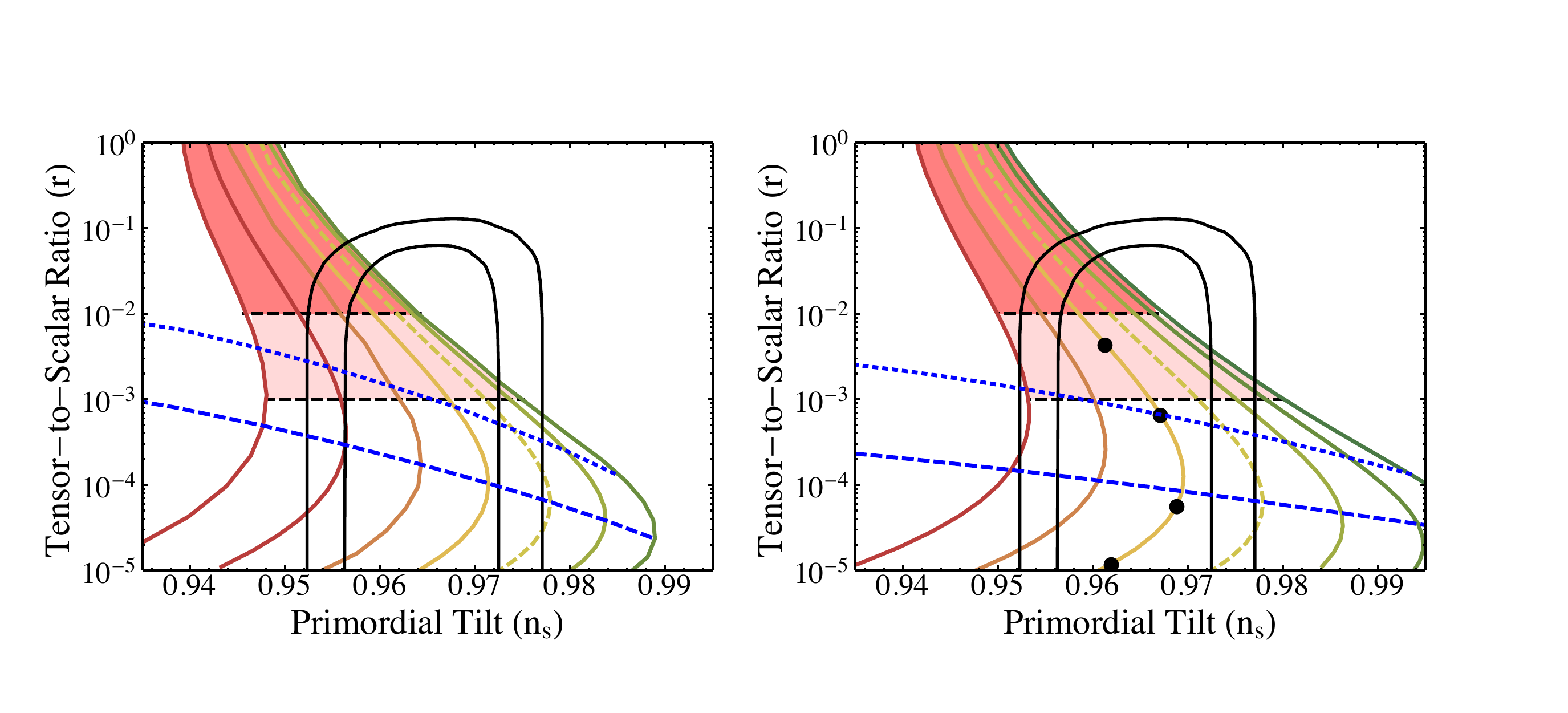}  
\caption{The tensor-to-scalar ratio, $r$, (at $k = 0.002 h/$Mpc$^{-1}$) and the spectral tilt, $n_s$(at $k = 0.05 h/$Mpc$^{-1}$) for models drawn from a grid of values for the parameters $g$,$f$,$\mu$,$\lambda$. The mode $k = 0.05 h/$Mpc$^{-1}$ is assumed to leave the horizon $60$ $e$-foldings before the end of inflation.  In both panels $Z_0$ is increased along each rainbow-colored curve (from top to bottom). In the left panel the different rainbow-colored curves have $\lambda=2600$ and $g/\mu^2$ varying from  $30$ to $60$ with a step of $5$ (left to right). The vertical dashed curve is common in the two panels.
 The diagonal blue dotted and blue dashed curves correspond to fixing $Z_0=0.035 $ and $Z_0=0.04 $ respectively and varying $g/\mu^2$ (left) or $\lambda$ (right). 
In the right panel the different rainbow-colored curves have $g/\mu^2=50$ and $\lambda$ varying from $2000$ to $3200$ with a step of $200$ (left to right). The diagonal blue dotted and dashed curves correspond to fixing $Z_0=0.035$ and $Z_0=0.04 $ respectively and varying $\lambda$. In both panels,  the shaded light red regions correspond to the 10\% and 1\% limits of the linear regime, as discussed in section\ \ref{sec:linvalid}.}
\label{fig:paramscan}
\end{figure}

\begin{figure}
\includegraphics[width=\textwidth]{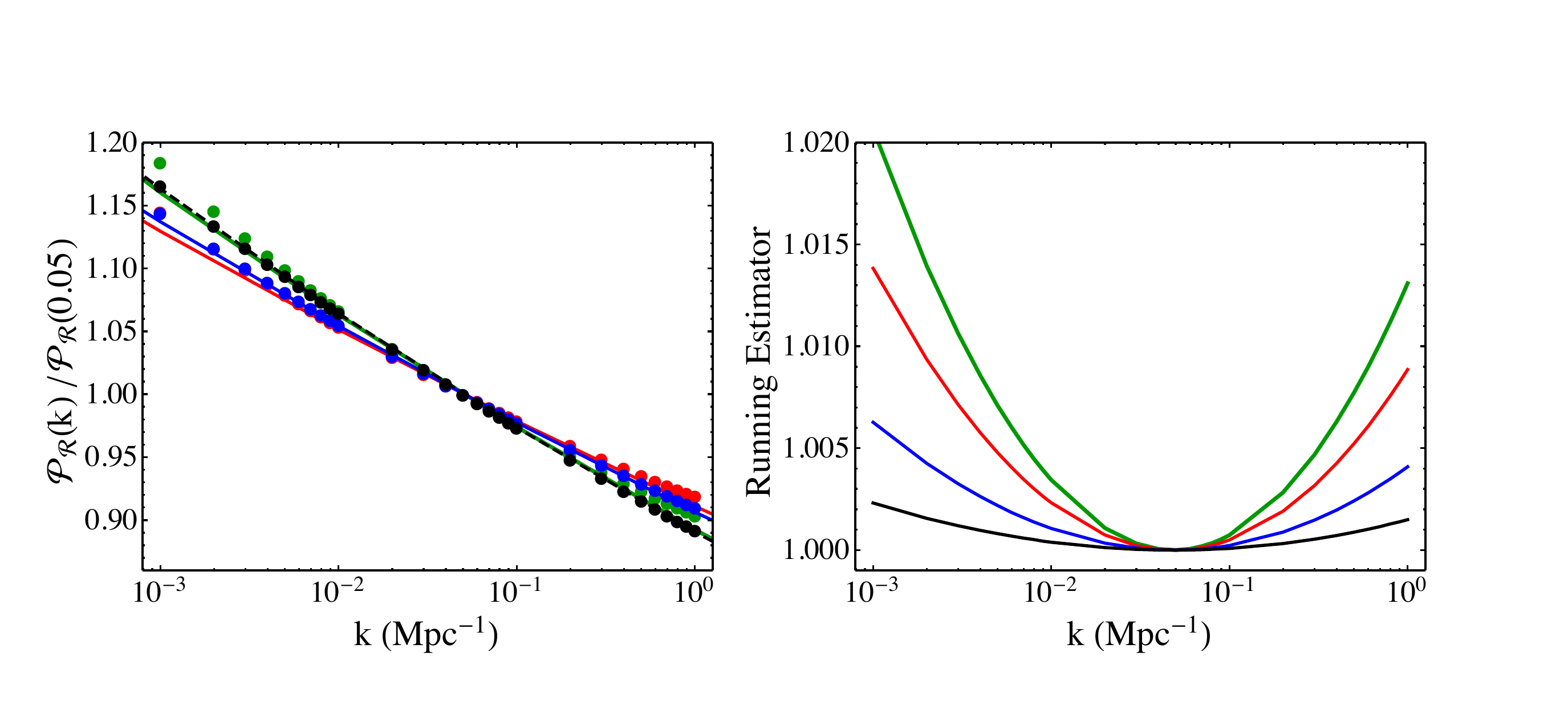}  
\caption{ 
Left: The scalar power spectrum amplitude normalized to unity at $k=0.05$ Mpc$^{-1}$ as a function of wavenumber for the parameters corresponding to the four dots of  figure \ref{fig:paramscan}. The green, red, blue and black lines (from smaller to larger $r$ respectively) correspond to power-law fits, $k^{n_s -1}$, while the dots show the results of numerical simulations. 
Right: The ratio of the normalized tensor amplitude (plotted on the left panel) to the form given in eq.\ 
\eqref{eq:fittingP}. This is a constant line in the case of zero running of the spectral index, hence it can be used as a visual estimator of the magnitude of the running $\alpha$. 
 }
\label{fig:running}
\end{figure}

\begin{figure}
\includegraphics[width=\textwidth]{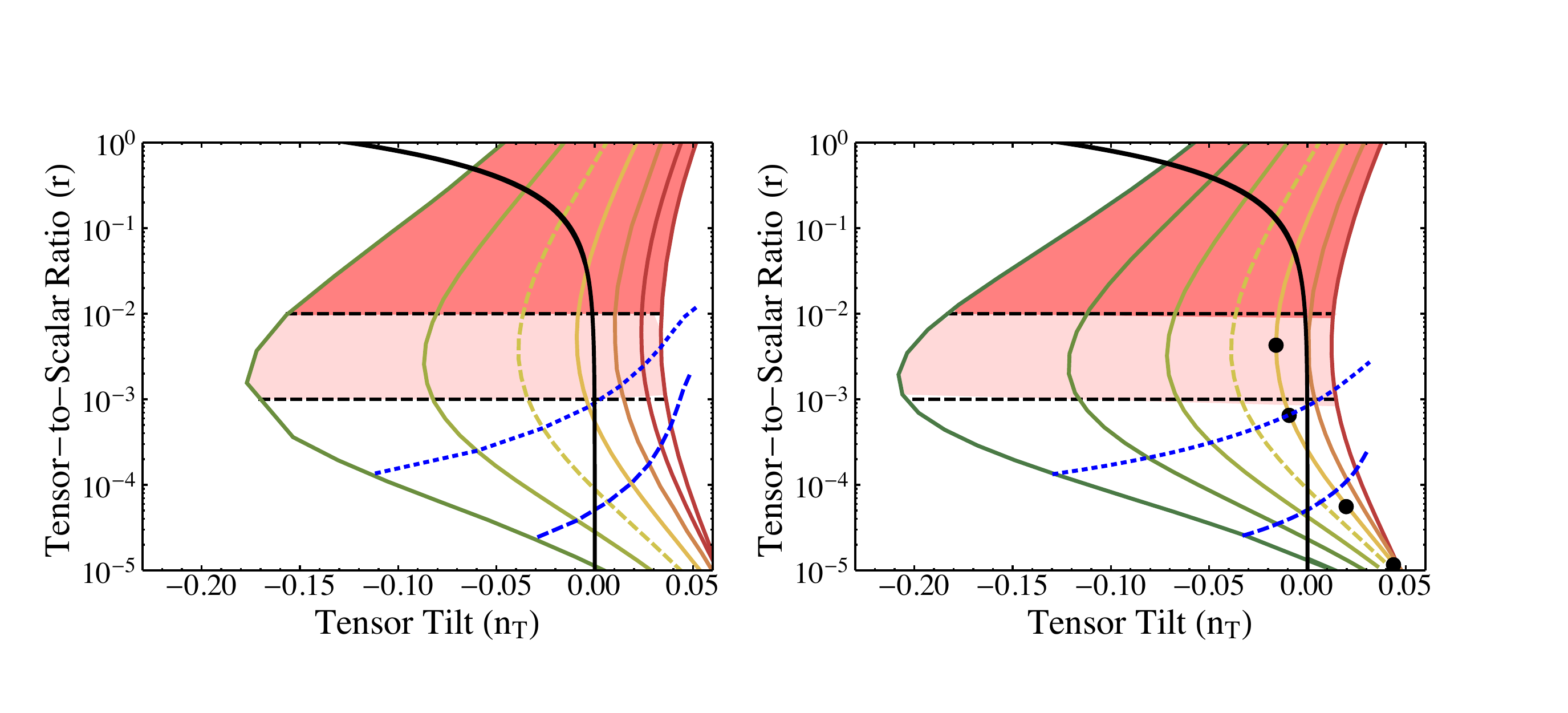}  
\caption{The tensor-to-scalar ratio, $r$, (at $k = 0.002 h/$Mpc$^{-1}$) as a function of the spectral tilt of the tensor modes, $n_T$ (at $k = 0.05 h/$Mpc$^{-1}$) for models drawn from a grid of values for the parameters $g$,$f$,$\mu$,$\lambda$. All parameters and color-coding are as in figure \ref{fig:paramscan}. The solid black line corresponds to the tensor consistency relation $r=-8n_T$. The color-coding follows the convention of Fig.\ \ref{fig:paramscan}. }
\label{fig:paramscanNt}
\end{figure}

The results of our numerical study are shown in figures \ref{fig:paramscan}, \ref{fig:running} and \ref{fig:paramscanNt}, and the parameters and the resulting spectral properties of some specific viable models are shown in table  \ref{tab:paramscan}. The figures were produced as follows. 
\begin{itemize}
\item  We construct a grid of values for each of the following model parameters: $g$, $\lambda$, $\mu$ and $Z_0$. The results are insensitive to the choice of the axion decay constant, $f$, and we fix $f=0.1$ for all runs. 

\item For each combination of parameters, we evolve the background equations to determine the axion location that corresponds to 
$N=60$ e-foldings prior to the end of inflation. We can then establish what conformal time corresponds to horizon exit for the ``pivot" momentum mode $k=0.05 \,h$/Mpc.

\item 
We evolve the perturbation equations for the scalars from an initial time corresponding to $x_{i}=-k \tau_{i} = 5\times10^4$ until a final time corresponding to $x_f = 10^{-3}$ with $k=0.05\,h$/Mpc and compute the curvature power spectrum and its tilt. We compute the amplitude of the tensor power spectrum by evolving the tensor equations for $k=0.002\,h$/Mpc and compare this with the amplitude of the scalar curvature perturbation at the same $k$ to evaluate the scalar-to-tensor ratio, $r$.

\end{itemize}
Since only the ``slow'' mode results in a significant final axion amplitude, we only initialize the computation in this mode, as described in section \ref{initialconditions}.

We begin by addressing the dependence of the observables ($r$, $n_s$) on the combination $g/\mu^2$. In section \ref{sec:backgroundparams}, we demonstrated that by varying $\mu$ and $g$, but keeping the ratio $g/\mu^2$ fixed, the number of $e$-folds of inflation does not change; it is natural to explore the dependence of the fluctuations on this combination. Fixing the ratio $g/\mu^2$, as well as $Z_0$ and $\lambda$, and varying the value of $\mu$, we found that both the tensor to scalar ratio, $r$, as well as the spectral index, $n_s-1$, change negligibly. However, the amplitude of scalar and tensor perturbations are proportional to $H^2 \sim \mu^4$, thus the overall amplitudes of the spectra can be tuned while $r$ and $n_s$ remain fixed. This means that once a combination of  parameters is found to give desired values $r$ and $n_s$,  ${\cal P}_{\cal R}$ can be adjusted by varying $\mu$ and $g$ while keeping  $g/\mu^2$ and the other parameters fixed.  We will thus only vary the combination $g/\mu^2$ in our scans. 

Figure \ref{fig:paramscan} shows the results of our parameter scan on the $n_s$-$r$ plane. In contrast to the case of Chromo-Natural Inflation \cite{Adshead:2013nka}, we are able to find regions of parameter space where the spectra are consistent with current data. We find that the Higgs VEV must satisfy $Z_0 \gtrsim 0.025$ in order for the model to fall within observational limits (linearity considerations further increase the bound on $Z_0$, as discussed in Section \ref{sec:linvalid}).  Increasing the value of the Higgs VEV generally reduces the tensor-to-scalar ratio $r$. This is due to the fact that both scalar modes and tensor modes are amplified, however, the scalar modes are more strongly amplified as $Z_0$ is increased. This behavior is evident from examination of figures  \ref{fig:scalarmodes} and \ref{fig:tensormodes}, note that increasing $M_{\hvev}$ here by a factor of two boosts the scalar spectrum by nearly three orders of magnitude, while the tensors are only boosted by approximately one and a half orders of magnitude.  In table\ \ref{tab:paramscan}, we present a series of specific scenarios from figure \ref{fig:paramscan}, highlighted using black dots. 

In order to keep ${\cal P_R} \simeq 2\times 10^{-9}$, we generally must reduce $\mu$ as we increase $Z_0$. For $Z_0 \approx 0.035$ the tensor-to-scalar ratio is $r = {\cal O}(10^{-3})$, which is small enough to be outside the region that is potentially observable in the immediate future $r \gtrsim {\cal O}(0.01)$. Further increasing $Z_0$ reduces the tensor to scalar ratio even more, making its detection impossible, even with next-generation experiments \cite{Abazajian:2013vfg}. 
 
 \begin{table}[h]
 \begin{center}
   \begin{tabular}{ | l | l | l | l | l | l | l | l | l | l | l | l |}
    \hline
    $\mu $  & $Z_0$  &  $\axion/f$  & $\psi$ & $m_\psi$ & $m_{Z_0}$ & $n_s$ & $r$ & $n_t$ & $\alpha \times 10^{4}$ \\ \hline
        $5.4 \times 10^{-4}$  & $0.031$  &  $2.42$ & $0.0185$ & $3.22$ & $5.39$ & $0.961 $ & $4.4 \times 10^{-3}$ & $-0.016$ & $2%\times 10^{-4} 
        $  \\ \hline
     $2.1 \times 10^{-4}$ &  $0.035$  &  $2.37$ & $0.0184$ & $2.99$ & $5.67$ & $0.967 $ & $6.6 \times 10^{-4}$ & $-0.009$ & $4%\times 10^{-4} 
     $ \\ \hline
     $4.5 \times 10^{-5}$ &  $0.041$  &  $2.28$& $0.0183$& $2.68$ & $5.99$ & $0.969 $ & $5.7 \times 10^{-5}$ & $0.020$ & $9$%\times 10^{-4}$ 
     \\ \hline
          $1.5 \times 10^{-5}$  & $0.045$  &  $2.22$& $0.0182$& $2.50$ & $6.17$ & $0.962 $ & $1.2 \times 10^{-6}$ & $0.044$ & $14$%\times 10^{-3}$ 
          \\ \hline
     \end{tabular}
     \end{center}
 \caption{Potential parameters and observables for the black dots shown in figure \ref{fig:paramscan}. The two parameters that remain constant are $g/\mu^2 = 50$ and $\lambda=2400$.} 
 \label{tab:paramscan}
\end{table}%

In order to provide a better understanding of the behavior of the inflationary spectrum, we plot the scalar spectrum for the parameters corresponding to the black dots in figure \ref{fig:paramscan} over three decades of wavenumbers $10^{-3} \le k < 1$ (in Mpc$^{-1}$)  in figure \ref{fig:running}. These can be fitted very well by a simple power law
\begin{align}
{\cal P}_{\cal R} \simeq \left ( k\over 0.05 \right ) ^{n_s-1}
\label{eq:fittingP}
\end{align}
where $n_s$ is the value presented in table\ \ref{tab:paramscan}.  If the power-law fit was exact, the running would vanish, since 
\begin{align}
\alpha \equiv {d n_s\over d \, \ln k} \simeq 0
\end{align}
for constant  $n_s$. In order to calculate the running we locally fit $\ln {\cal P}_{\cal R}$ as a function of $\ln k$ using a second order polynomial around $k=0.05$. The resulting running of the primordial spectral index is shown in table\ \ref{tab:paramscan}. From this small subset of parameters, we observe that the running of the tilt is positive, in contrast to simple single-field models (see, for example \cite{Adshead:2010mc}) and increases as the tensor-to-scalar ratio is decreased.   The constraints from Planck \cite{Ade:2015lrj} are 
$\alpha = -0.0084 \pm 0.0082$.
This result can be lowered to $ \alpha = -0.0033 \pm 0.0074$ if the high$-l$ polarization and CMB lensing data is included. Thus for values of the tensor-to-scalar ratio larger than $r\simeq10^{-6}$ our model is consistent with the Planck data.  The trend of table\ \ref{tab:paramscan} indicates that for $r<10^{-6}$ the observables can be in conflict with the data due to the large positive running of the spectral tilt, $\alpha$. Between this and the linearity considerations (discussed in section \ref{sec:linvalid}) this model provides a viable band of observables in accordance with present observational data.

Since the value of the axion decay constant $f$ does not affect the observables, the inflaton can be arranged to have arbitrarily sub-Planckian field excursions regardless of the tensor-to-scalar ratio, in violation of simple formulations of the Lyth bound \cite{Lyth:1996im}. However, a more meaningful comparison is between the Hubble rate during inflation and the amplitude of primordial gravitational waves produced --  in the simple single-field inflationary scenario, the gravitational wave spectrum depends only on the Hubble rate during inflation.  Thus, in standard single field inflation, a measurement of the gravitational wave power spectrum (via the B-mode of the CMB) is a direct measurement of the energy scale of inflation, which can be inferred from
\begin{align}\label{eqn:SSFI}
\frac{H^{2}_{\rm Inf}}{M_{\rm Pl}^2} = \frac{\pi^2}{2} \Delta_{\mathcal{R}}^2 r .
\end{align}
 The model presented here explicitly breaks this connection by exponentially enhancing the amplitude of the tensor fluctuations. Therefore, in this scenario, a measurement of the tensor-to-scalar ratio does not directly measure the Hubble rate during inflation.  In particular, for $r = 4.4 \times 10^{-3}$, the inferred value of $H$ from eq.\ \eqref{eqn:SSFI} is $H^2 \approx 5.3 \times 10^{-11}M_{\rm Pl}^2$, while the actual value computed from table \ref{tab:paramscan} is somewhat lower: $H^2 \approx 7.1 \times 10^{-15} M_{\rm Pl}^2$. %Consequently, the inferred inflationary energy scale is $\Lambda_{I} = (3M_{\rm Pl}^2H^2)^{1/4}  \sim 0.004 M_{\rm Pl}$, while inflation actually occurs at the scale $\Lambda_{I} = (3M_{\rm Pl}^2H^2)^{1/4} \sim  0.0004 M_{\rm Pl}$.

Furthermore, in standard single field scenarios, the tensor-to-scalar ratio and the tilt of the tensor spectrum obey a consistency condition
\begin{align}
 n_T = -\frac{r}{8},
\end{align}
which guarantees a red-tilted spectrum ($n_T < 0$). In figure \ref{fig:paramscanNt} we display the relationship between $r$ and $n_T$ in Higgsed Chromo-Natural Inflation.  Note that in the regions of parameter space preferred by the Planck data (the yellow - orange curves) the gravitational wave spectra can be either blue- or red-tilted.

\subsection{Validity of the linear theory}\label{sec:linvalid}

We end this section with some comments about the linearity of the perturbations. The initial motivation for adding mass to the theory was to try and tame the instability in the spin-2 sector of the gauge field modes at large $\mpsi$ in order to yield scalar fluctuations with acceptable scale dependence. It is evident from eq.\ (\ref{eqn:tensorinstability}) that this expectation is not borne out. The instability persists and, for moderate values of the gauge field mass, the spin-2 modes of the gauge field fluctuations still attain significant amplitudes. Further, the additional Goldstone mode dynamics also lead to the amplification of the scalar and vector parts of the gauge field.

It thus behooves us to determine if our linearized treatment is under control -- we are working at linear order in perturbation theory, and making use of a scalar-vector-tensor decomposition of the fluctuations. It is thus natural ask at what point our assumptions become invalidated by these growing field fluctuations? We do not attempt to answer this question carefully here, but provide a quick estimate of when non-linearity could become important. 

From figures \ref{fig:scalarmodes}, \ref{fig:tensormodes} and  \ref{fig:vectormodes}, the gauge field modes undergo a period of amplification that begins near $-k\tau_* \sim M_{\hvev}$ and continues until near $-k\tau \sim 1$, as their wavelengths become equal to the Hubble rate, and they freeze out. However, we need to determine if these modes could be sourcing non-linearities. 

In order that our linearized theory is under control, we require that the background field fluctuations, $\delta A_\mu$ are smaller than the classical field value itself $\bar{A}_\mu = (0, a\psi \delta^a_i J_a)$,
\begin{align}
\frac{|\dA_{\mu}|}{|\bar{A}_{\mu}|} \ll 1.
\end{align}
We can estimate the size of our fluctuations relative to the background by computing the root-mean-square (rms) value of the field fluctuations
\begin{align}\label{eqn:rms}
|\dA| = \sqrt{\langle \dA^2 \rangle} = \sqrt{\int \frac{d^3 k}{(2\pi)^3}|\dA_k|^2} \sim  \frac{(M_{Z_0}aH)}{2\pi} |\sqrt{2k}\,\dA_k(k\tau_*) |
\end{align}
where we cut the integral off at the peak of the amplification, which is observed to be near $k/aH \sim M_{Z_0}$.  Comparing eq.\ \eqref{eqn:rms} to the classical background gauge field shows that our theory is we described by the linear approximation provided that
\begin{align}
\frac{(M_{Z_0}H)}{2\pi} |\sqrt{2k}\,\dA_k(k\tau_*) | \ll  \psi.
\end{align}
Therefore, provided that the gauge field amplification remains much smaller than approximately $H^{-1}$, our linear theory should remain well under control. Note that this can also be rewritten
\begin{align}
\frac{|\sqrt{2k}\,\dA_k(k\tau_*) |}{2\pi} \ll \frac{1}{g}\frac{m_{\psi}}{M_{Z_0}}.
\label{eq:linearitycondition}
\end{align}
We now can impose these limits on our parameter scan. In the region shown in figure \ref{fig:paramscan}, that is for $0.1 >  r >  10^{-4}$, the typical maximum value that the gauge-tensor attains is $|\sqrt{2k}\,\dA_k(k\tau_*) | \simeq 10^{4}$. We then indicate in red bands the regions where  eq.\ \eqref{eq:linearitycondition} is worse than $10\%$ and $1\%$ respectively for the value of the gauge coupling chosen so that the scalar amplitude matches the value measure by Planck.  For $r \lesssim 10^{-3}$, the system appears to be well within the linear regime according to the estimate in eq.\ \eqref{eq:linearitycondition}. Note that large values of the tensor-to-scalar ratio $r \gtrsim 0.01$ are likely to be in the non-linear regime. In this regime, the non-linear effects of the gauge fields can not be neglected. The situation is less clear for $10^{-3} \lesssim r \lesssim 10^{-2} $.

While it seems clear that there is a well defined region of observables when our linear perturbation theory is under control, it is less clear if the resulting fluctuations are close to Gaussian. We leave the study of non-Gaussianities in this theory to future work.

%%%%%%%%%%%%%%%%%%%%%%%%%%%%%%%%%%%%
%%%%%%%%%%%%%%%%%%%%%%%%%%%%%%%%%%%%
%%
\section{Higgsed Gauge-flation}\label{sec:HGF}
%%
%%%%%%%%%%%%%%%%%%%%%%%%%%%%%%%%%%%%
%%%%%%%%%%%%%%%%%%%%%%%%%%%%%%%%%%%%

In specific cases where the axion is close to the bottom of its potential throughout the entirety of the inflationary evolution, it can be integrated out to give a related model of inflation, Gauge-flation \cite{Maleknejad:2011jw,Maleknejad:2011sq} (see \cite{Adshead:2012qe, SheikhJabbari:2012qf}).  Gauge-flation is described by the action
\begin{align}\label{eqn:GFact}
\mathcal{S} =  \int d^{4}x\sqrt{-g}\Bigg[  \frac{1}{2}R -\frac{1}{2}\tr\[F_{\mu\nu}F^{\mu\nu}\]+\frac{\kappa}{96}\tr\[ F\wedge F \]^2\Bigg] ~,
\end{align}
and thus is an example of a model of inflation where the accelerated expansion is not driven by a scalar degree of freedom. Unfortunately, as mentioned above, at the level of the fluctuations Gauge-flation does not result in a viable cosmology for the same reasons as Chromo-Natural Inflation \cite{Adshead:2013nka, Namba:2013kia}. This is unsurprising given that the models are very closely related; Gaugflation may be put in a form suitable for applying the analysis of Chromo-Natural Inflation with the introduction of a pseudo-scalar auxiliary field $\axion$
\begin{align}\nn\label{eqn:CNIform}
\mathcal{S} =  \int d^{4}x\sqrt{-g}\Bigg[  \frac{1}{2}R -\frac{1}{2}\mu^4\( \frac{\axion}{f}\)^2-\frac{1}{2}\tr\[F_{\mu\nu}F^{\mu\nu}\]-\frac{\lambda}{4 f}\axion\tr\[ F\wedge F \]\Bigg] ~.
\end{align}
Integrating out this auxiliary field yields the Gauge-flation action with the identification of parameters
\begin{align}
\kappa = 3\frac{\lambda^2}{\mu^4} ~.
\end{align}
The introduction of a Higgs sector to this theory would give ``Higgsed Gauge-flation"\footnote{While this work was in progress similar ideas were proposed in ref.\ \cite{Nieto:2016gnp}}
\begin{align}\label{eqn:GFact}
\mathcal{S} =  \int d^{4}x\sqrt{-g}\Bigg[  \frac{1}{2}R -\frac{1}{2}\tr\[F_{\mu\nu}F^{\mu\nu}\]+\frac{\kappa}{96}\tr\[ F\wedge F \]^2  -g^2\hvev^2\tr\[ A_{\mu} - \frac{i}{g}U^{-1}\partial_\mu U\]^2\Bigg] ~,
\end{align}
where as above, $U = \exp(i g \xi)$. It would be interesting to check to see if this theory can provide viable inflationary scenarios. We leave detailed investigation of Higgsed Gauge-flation to future study.

%%%%%%%%%%%%%%%%%%%%%%%%%%%%%%%%%%%%
%%%%%%%%%%%%%%%%%%%%%%%%%%%%%%%%%%%%
%%
\section{Conclusions}\label{sec:conclusions}
%%
%%%%%%%%%%%%%%%%%%%%%%%%%%%%%%%%%%%%
%%%%%%%%%%%%%%%%%%%%%%%%%%%%%%%%%%%%

In this work we have shown that Chromo-Natural Inflation can be potentially made compatible with existing limits from Planck data by introducing an additional mass term for the gauge field fluctuations. In this work, we assume that the symmetry is spontaneously broken by a Higgs sector and the resulting Higgs boson is much heavier than the Hubble scale, and is thus irrelevant. We therefore work with the theory in the Stueckelberg form.

While the addition of the Stueckelberg symmetry breaking sector was initially motivated to provide a stabilization mechanism for the spin-2 modes of the gauge field by giving it an additional mass, this does not in fact  happen. The reason is that such a mass term also contributes to the equations of motion at the background level, leading to larger values of the axion velocity which sources the tensor instability.  However, the Goldstone modes contribute additional scalar and vector degrees of freedom at the level of the fluctuations. The interaction of the additional scalar degree of freedom boosts the curvature fluctuation relative to the tensor fluctuations. This consequently lowers the tensor-to-scalar ratio into the region allowed by BICEP, the Keck Array, and the Planck satellite \cite{Ade:2015tva, Ade:2015lrj}. 

Observable gravitational waves ($r \gtrsim 10^{-3}$) may be produced in this model, despite inflation occurring below the GUT scale, and all fields evolving over sub-Planckian distances in field space.  The model therefore violates some formulations of the Lyth bound. The gravitational waves in this model predominantly arise from linear mixing with the gauge field fluctuations. These gauge field modes are enhanced by their interactions with the rolling axion and subsequently oscillate into gravitational waves. The form of the gravitational wave spectra produced in this model is therefore significantly altered from the usual form assumed in formulations of the Lyth bound.  In contrast to standard inflationary scenarios which uniformly predict red tilted gravitational wave spectra (see, however, \cite{Baumann:2015xxa}), these gravitational waves can have either red- or blue-tilted spectra on CMB scales. Furthermore, these gravitational waves have the distinct characteristic that they are chirally polarized and, to a very good approximation, consist only of a single helicity. Unfortunately, it seems that future CMB experiments will be unable to distinguish between unpolarized and chirally polarized gravitational waves \cite{Gerbino:2016mqb}. 

The equations of motion for the field fluctuations that result in this system are complicated, but are fairly simple to solve numerically. Of the four normal modes of the system, only the mode with the smallest frequency (the slow, or magnetic drift mode) results in fluctuations which attain significant superhorizon amplitude.  At first glance, one may worry that the presence of multiple large-amplitude scalar modes on superhorizon scales may lead to pathological effects, such as isocurvature or entropy fluctuations which cause the curvature perturbation to evolve. However,  we have shown that  entropy or isocurvature fluctuations are suppressed relative to adiabatic curvature fluctuations, and contribute only at the sub-percent level. The dominant contribution to the perturbed stress-energy tensor is due to the axion's fluctuations along its potential, and since this term dominates the evolution of the background, the non-adiabatic pressure is small.

We have demonstrated that the parameters of the theory can be chosen to produce fluctuations that, near horizon crossing, match the required amplitude and tilt of the scalar spectrum as determined by the Planck satellite \cite{Ade:2015lrj}. While we have neglected the contributions of the metric fluctuations (in the form of the perturbed lapse and shift) in this work, we expect that including these will alter our results at the level of slow roll corrections. The fluctuations in the gauge field and axion, as well as the Goldstone modes, depend exponentially on the Higgs VEV, which makes some level of fine-tuning necessary in order to match observations.  

For parameters leading to large values of the tensor-to-scalar ratio, $r > 0.1$, our estimates suggest that the linear approximation used in deriving the equations of motion for the fluctuations likely fails. For $r\gtrsim 10^{-2}$ and $r\gtrsim 10^{-3}$, we estimates that the linearity fails on the level of $10\%$ and $1\%$, respectively. 

A potentially significant restriction on this model comes from the running of the spectral index. While this running is negligible for parameters leading to $r\gtrsim 10^{-3}$, as $r$ decreases we have found that the running increases. Furthermore, the running of the tilt in this model is positive, in contrast to many single field models that predict negative running at the $\mathcal{O}\bigl((n_s \!-\!1)^2\bigr)$ level (see, e.g.\ \cite{Adshead:2010mc}). This is also in contrast to the slight preference for negative running observed in the CMB data \cite{Ade:2015lrj}. For $r\simeq 10^{-5}$ the running of the spectral index remains within the observational bounds set by the Planck mission, however, for very low values of $r$ this model will likely be ruled out.

Throughout this work, we have neglected the contribution of metric fluctuations as well as slow-roll corrections to the equations of motion. For an initial investigation this is most likely a good approximation, at least until after horizon crossing where we evaluate the spectra. However,  computation of the full evolution of the modes outside the horizon requires a more careful analysis that includes the contributions from the gravitational constraints and the slow-roll corrections due to the evolution of the background. We leave this, as well as detailed investigations of non-Gaussianity  to future work.  Finally, given that adding Higgs sector to Chromo-Natural inflation potentially yields viable cosmologies, it would be interesting to check whether the related model of Gauge-flation can be made viable in the same fashion.

 %%%%%%%%%%%%%%%%%%%%%%%%%%%%%%%%%%%%
%%%%%%%%%%%%%%%%%%%%%%%%%%%%%%%%%%%%
%%%%%%%%%%%%%%%%%%%%%%%%%%%%%%%%%%%%
%\vskip 2cm

{\bf Acknowledgements:}
This work was supported in part by DOE grants DE-FG02-90ER-40560, DE-SC0009924, DE-SC0015655 and by the Kavli Institute for Cosmological Physics at the University of Chicago through grants NSF PHY-1125897 and an endowment from the Kavli Foundation and its founder Fred Kavli. P.A. gratefully acknowledges support from a Starting Grant of the European Research Council (ERC STG grant 279617), and the hospitality of DAMTP and the University of Cambridge where some of this work was completed.
EIS gratefully acknowledges support from a Fortner Fellowship at the University of Illinois at Urbana-Champaign.

\appendix

\section{Conventions}\label{App:conventions}

We adopt the conventions of  Peskin and Schroeder~\cite{Peskin:1995ev} for the action of the gauge field. In particular, the field-strength tensor and covariant derivative are defined as\footnote{Note that this is opposite to \cite{Adshead:2012kp, Maleknejad:2011jw, Maleknejad:2011sq}, where the opposite sign for the covariant derivative was used.}
\begin{align}
F_{\mu\nu} = \frac{1}{-ig}\[D_{\mu}, D_\nu\], \quad D_{\mu} = \partial_{\mu} - igA_{\mu},
\end{align}
where $g$ is the gauge field coupling, not to be confused with the determinant of the spacetime metric. We normalize the trace over the SU(N) matrices, which we denote $J_a$, so that
\begin{align}
\tr\[J_a J_b\] = \frac{1}{2}\delta_{ab}.
\end{align}
Our convention for the antisymmetric tensor is
\begin{align}
\epsilon^{0123} = \frac{1}{\sqrt{-g}}.
\end{align}
while our spacetime metric signature is $(-,+,+,+)$.  Here and throughout, Greek letters  denote spacetime indices, Roman letters from the start of the alphabet denote gauge indices and Roman letters from the middle of the alphabet denote spatial indices.

We work with conformal time, which we define to be a negative quantity during inflation
\begin{align}
\tau = \int^{t}_{0}\frac{dt}{a(t)},
\end{align}
and make use of the near de Sitter expansion to write
\begin{align}
a \approx -\frac{1}{H \tau}.
\end{align}
When we are dealing with fluctuations of the fields, we work in Fourier space where our convention is
\begin{align}
A({\bf x}) = \int \frac{d^3 k}{(2\pi)^3}A_{\bf k}e^{-i {\bf k}\cdot{\bf x}},
\end{align}
so that we replace spatial derivatives with
\begin{align}
\partial_i A \to -i k_i A_{\bf k}.
\end{align}
We make extensive use of the fact that the fields satisfy a reality condition, which implies
\begin{align}
A_{-{\bf k}} = \bar{A}_{\bf k}.
\end{align}
It  will often prove useful to work with the dimensionless time variable 
\begin{align}
x = -k\tau,
\end{align}
where $k$ is the Fourier space wavenumber. When we match to observations, we take $k$ to have cosmological units, $h$/Mpc, which also fixes the units for $\tau$. Where necessary, we match physical length scales to inflationary scales by choosing the scale $k = 0.05$ Mpc$^{-1}$ to leave the horizon 60 $e$-folds before the end of inflation. Throughout we  denote derivatives with respect to cosmic time by an overdot ($\,\dot{}\,$), primes ($\,'\,$) denote derivatives with respect to $x$, while derivatives with respect to conformal time are kept explicit ($\partial_\tau$). Our symmetrization and antisymmetrization conventions throughout are
\begin{align}
Z_{[ij]} = & \frac{1}{2}(Z_{ij} - Z_{ji}), \quad Z_{(ij)} =  \frac{1}{2}(Z_{ij} + Z_{ji}).
\end{align}

\section{A specific realization: Adjoint Higgs model}\label{app:specificmodel}

In the main text we worked only with the Higgs action in Stueckelberg form where all fluctuations are taken to be along the vacuum manifold. In this limit all models where a Higgs is introduced must be identical. However, away from this limit there is considerable freedom.

In general, a Higgs field $Z$ with a general potential
\begin{align}
\mathcal{L}_{Z} = \sqrt{-g}\[ -\frac{g^{\mu\nu}}{2}D_{\mu}ZD_{\nu}Z^{\dagger} - V(Z)\]
\end{align}
has a stress tensor
\begin{align}
T_{\mu\nu} = D_{\mu}Z D_{\nu}Z^{\dagger}  - g_{\mu\nu}\mathcal{L}_{Z}.
\end{align}
Note that, due to the background gauge field,  if the background value of the Higgs field is allowed to evolved with some $\dot Z \neq 0$, there is a non-zero momentum flux
\begin{align}
T_{0i} = \partial_0 Z A_i Z^{\dagger} - g_{0i}\mathcal{L}_{Z}.
\end{align}
It thus initially appears that unless the Higgs is completely fixed on its vacuum manifold then the resulting stress-energy tensor is in fact inconsistent with the symmetries of FRW spacetime. However, in this section we introduce an explicit model that is compatible with the symmetries of FRW regardless of the evolution of the Higgs to allay these concerns. We first note that, if we choose a triplet of Higgs fields in the adjoint representation of SU(2),
\begin{align}
Z_A = Z^a_A J_a,
\end{align}
where $A \in \{1, 2, 3\}$ is a field index, then we can choose the expectation value to be of the form
\begin{align}
Z_A = Z_0(t)\delta^a_A J_a.
\end{align}
For this field configuration, one has an additional residual SO(3) symmetry with which to protect the background spacetime and in this case it is straightforward to see that $T_{0i}$ vanishes independently of $\dot Z$. 

Generically, such a (matrix valued) scalar field has a potential of the form
\begin{align}
V(Z) = \tr\[-\frac{\Upsilon}{4}\[Z_A, Z_B\]^2 + \frac{i \kappa}{3} \epsilon_{ABC} \[Z_A, Z_B\]Z_C + \frac{m^2}{2}Z_A Z_A\].
\end{align}
Note that, by choosing 
\begin{align}
\kappa = \frac{3}{2}\Upsilon \beta, \quad m^2 = \Upsilon \beta^2
\end{align}
for the above configuration, the potential is put in the symmetry breaking form
\begin{align}
V(Z) = \frac{3}{2}\Upsilon Z_0^2(Z_0 - \beta)^2.
\end{align}
So that for large values of $\Upsilon$, the classical Higgs is confined to $Z_0 = \beta$. Moreover, it is straightforward to show that quadratic fluctuations about the minimum at $Z_0 = \beta$ \cite{Ashoorioon:2009wa}
\begin{align}
V^{(2)} = \frac{\Upsilon \beta^2}{2}\[\omega^2+2\omega+1\]\delta Z^A_a \delta Z^A_a,
\end{align}
where $\omega$ are the Eigenvalues of $\tilde{\Omega}_a = \omega \delta Z_a$ where
\begin{align}
\tilde{\Omega}_a = i\epsilon_{abc}\[\delta Z_b , J_c\].
\end{align}
We can further decompose
\begin{align}
Z^A = Z_0 \delta^A_a J_a+ \delta Z^A_a J_a = \exp\[-i{\rm adj }(\xi)\]\((Z_0+\delta Z)\delta^A_a +K^A_a \)J_a,
\end{align}
where $K^A_a$ is a traceless symmetric matrix. Under this decomposition the modes $\delta Z$ and $K^A_a$ have eigenvalues $\omega = -2$ and $\omega = 1$ respectively, and are thus modes with masses $\Upsilon \beta^2/2$ and $2 \Upsilon \beta^2$. The $\xi$ modes have eigenvalue $-1$, and are thus the massless Goldstone bosons corresponding to fluctuations along the vacuum manifold. 

In the main text we have worked in the limit where $\Upsilon \beta^2 \gg H^2 $ so that the fluctuations of these massive modes are irrelevant for cosmology. However, we note that one could introduce symmetry breaking patterns to the background spacetime by allowing for evolving Higgs vacua in, for example, the fundamental representation. We leave the study of these effects to future work.

\section{Details of the scalar action}\label{app:scalargore}

In this appendix we present the details of the matrices for the canonically normalized scalar modes $\Delta = (\phihat, \zhat, \ada, \hat{H})$ in their full gore. The anti-Hermitian  $4\times4$ matrix $K$ from section \ref{sec:scalarmodes} has non-zero components
\begin{align}
%K_{11} & =  K_{22} = K_{33} = K_{44} = K_{12} = K_{21} = K_{14} = K_{41}= 0\\
%
K_{13} = & -K_{31} \simeq -\frac{1}{2} \frac{\lambda \psi}{f}\frac{\mpsi}{x}\frac{1}{\sqrt{1+\frac{x^2}{2\mpsi^2}}}\\ 
K_{14} =& -K_{41} \simeq    - \frac{\sqrt{2}M_{\hvev} \mpsi }{(2\mpsi^2+x^2)}\frac{1}{\sqrt{M_{\hvev}^2+2\mpsi^2+x^2}},\\
K_{23} = & -K_{32} \simeq  \frac{1}{\sqrt{2}}\frac{\lambda\psi}{f}\frac{\mpsi}{x}\\
K_{34} = & -K_{43} \simeq  -\frac{1}{2}\frac{\lambda \psi}{f}\frac{M_{\hvev}\mpsi }{\sqrt{(2\mpsi^2+x^2)\(M_{\hvev}^2+2\mpsi^2+x^2\)}}
\end{align}
with all remaining components zero. The symmetric frequency matrix $\Omega^2$ has entries
\begin{align}\nn
\Omega^2_{11} \simeq & 1+ \frac{8M_{\hvev}^2\mpsi^2 }{\(M_{\hvev}^2+2  \mpsi^2+ x^2\)\left(2
    \mpsi^2+x^2\right)^2}+\frac{\mpsi^2}{\left(2
    \mpsi^2+x^2\right)}+\frac{3}{\left(2
    \mpsi^2+x^2\right)^2}\\ &+\frac{M_{\hvev}^2+2\mpsi^2}{ x^2} -\frac{\mpsi}{2\mpsi^2+x^2}\frac{\lambda\dot{\axion}}{Hf},\\
\Omega^2_{12} \simeq & -\frac{\sqrt{2 \mpsi^2+x^2}}{ x^2}\(2  \mpsi-\frac{\lambda\dot\axion}{Hf}\),\\
\Omega^2_{13} \simeq & -\frac{\lambda\psi}{f}\frac{\sqrt{2} \left(M_{\hvev}^2 \left( \mpsi^2+x^2\right) \left(3
    \mpsi^2+x^2\right)+\left(6  \mpsi^6+7  \mpsi^4 x^2+4  \mpsi^2
   x^4+x^6\right)\right)}{x^2 \left(2  \mpsi^2+x^2\right)^{3/2} \left(M_{\hvev}^2+2  \mpsi^2+
   x^2\right)},
\end{align}
 \begin{align}
 \Omega^2_{14} \simeq & -\frac{\sqrt{2}M_{\hvev}  \mpsi \left(M_{\hvev}^4+2 M_{\hvev}^2 \left(2  \mpsi^2+x^2+1\right)+ \left(2
    \mpsi^2+x^2\right)^2+4  \mpsi^2+5 x^2\right)}{\left(2  \mpsi^2
   x+x^3\right) \left(M_{\hvev}^2+2  \mpsi^2+x^2\right)^{3/2}},\\
 \Omega^2_{22} \simeq & 1-\frac{  \mpsi }{ x^2}\frac{\lambda\dot\axion}{Hf}+\frac{ 1}{x^2}\left(M_{\hvev}^2+  4\mpsi^2\right),\\
 \Omega^2_{23} \simeq & \frac{3}{\sqrt{2}}\frac{\lambda\psi}{f}\frac{\mpsi}{x^2},\quad \Omega^2_{24} = 0,\\
 %
 %\Omega^2_{24} \simeq & 0\\
 %
 \Omega^2_{33} \simeq &1 -\(\frac{2}{x^2}-\frac{V''}{H^2 x^2}-\frac{1}{f^2}\frac{\lambda^2\psi^2 \mpsi^2}{M_{\hvev}^2+2
    \mpsi^2+x^2}\),\\
  \Omega^2_{34} \simeq &-\frac{\lambda\psi}{f} \frac{ \mpsi M_{\hvev}\left(M_{\hvev}^2 \left(4  \mpsi^2+x^2\right)+4
   \left( \mpsi^2+x^2\right) \left(2  \mpsi^2+x^2\right)\right)}{2 x \left(\left(2
    \mpsi^2+x^2\right) \left(M_{\hvev}^2 +2  \mpsi^2+ x^2\right)\right)^{3/2}},\\\nn
  \Omega^2_{44} \simeq &1-\frac{2}{x^2}- \frac{2 M_{\hvev}^2  \mpsi^2 \left(M_{\hvev}^2+8  \mpsi^2+4 x^2\right)}{\left(2  \mpsi^2+x^2\right)^2
   \left(M_{\hvev}^2+2  \mpsi^2+x^2\right)^2}+\frac{3 M_{\hvev}^2}{\left(M_{\hvev}^2+2
    \mpsi^2+x^2\right)^2}+\frac{M_{\hvev}^2+2}{2  \mpsi^2+x^2}\\ & -\frac{2}{M_{\hvev}^2+2
    \mpsi^2+x^2}
 \end{align}
In the above expressions the `$\simeq$' refers to the fact that we have worked in the slow roll limit, dropping the variation of the background.

\section{Vector fluctuations}\label{App:vectors}
\begin{figure}[t!]
\centering
\includegraphics[width=\textwidth]{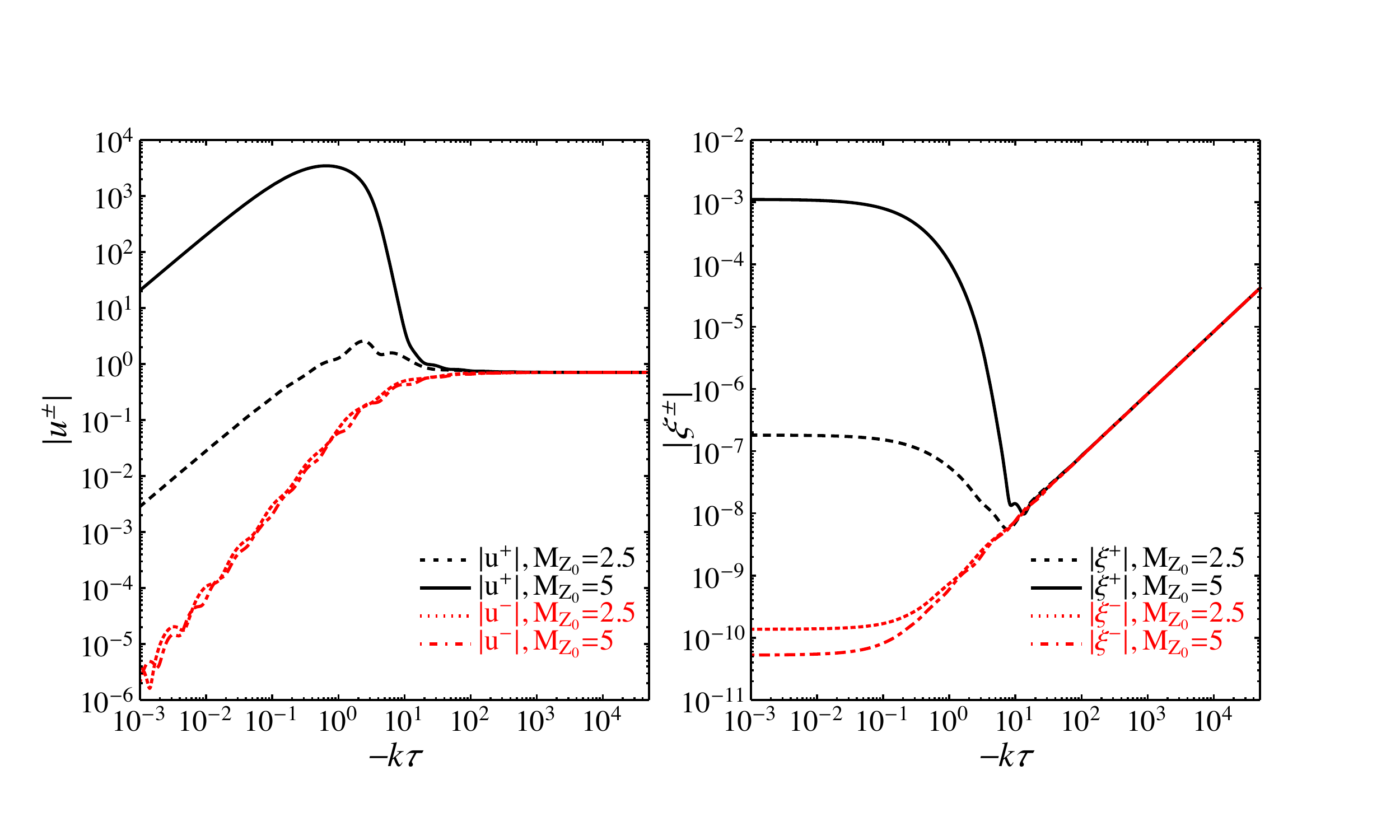}
\caption{
Evolution of vector fluctuations in Higgsed Chromo-Natural inflation. The addition of the Higgs leads to a temporary exponential instability in the vector modes that begins near $k\tau \sim M_{Z_0}$. At late times, the gauge vectors decay as $1/a$, while the Higgs vector freezes out. The values of the other parameters here are chosen to be $\mu  = 8 \times 10^{-5}$, $g =  
 1.28\times 10^{-7}$, $\lambda \psi/f = 3920$, $\mpsi \approx 2.5$,  $H \approx 1.2 \times 10^{-9}$, $\psi \approx 0.022$.
 }
\label{fig:vectormodes}
\end{figure}

For completeness, we now examine the vector degrees of freedom in the theory. We work in the basis,
\begin{align}
u^{\pm} = & \frac{1}{\sqrt{2}}(\chi^1 \pm i \chi^2), \quad v^{\pm} =  \frac{1}{\sqrt{2}}\(t_{31}\pm i t_{32}\),\quad
\dHiggs^{\pm} = \frac{gZ_0}{\sqrt{2}}\( \dHiggs^1 \pm i \dHiggs^2\).
\end{align}
In this basis, working in the slow-roll limit, and using the variable $x = -k\tau$, we have for the vector degrees of freedom of the gauge fields from the Yang-Mills and Chern-Simons actions
\begin{align}\nn\label{eqn:helvecaction2}
\frac{\delta^2 \mathcal{L}_{V}}{k^2} = & 
\partial_{x}v_{\pm}\partial_{x}\bar{v}_{\pm}+\partial_x  u_\pm \partial_x \bar{ u}_\pm-  v_{\pm}\bar{ v}_{\pm}  -  u_\pm \bar{ u}_\pm \\\nn
&+\(2\mpsi^2-8 \frac{\mpsi^2}{x^2\pm 2\mpsi x+2\mpsi^2+  M_{\hvev}^2}\)\frac{ u_\pm\bar{ u}_\pm}{x^2}+\frac{\mpsi}{x} \( v_{\pm}\bar{ v}_{\pm}\mp  u_\pm\bar{ u}_\pm\) \big]\\\nn & 
-(\mpsi^2+ M_{\hvev}^2)( v_{\pm}\bar{ v}_{\pm}+  u_\pm\bar u_\pm )-\frac{\mpsi}{x^2}\(\frac{\lambda}{f}\frac{\dot\axion}{H}-\mpsi \) \( v_{\pm}\bar{ v}_{\pm}-  u_\pm\bar u_\pm\)\\
%
%& +\frac{\lambda}{2f}\frac{\dot\axion}{H}\big[\pm   v_{+}\bar{ v}_{+}+i\( v_+ \bar{ u}_{+}+ v_-\bar{ u}_- -{\rm c.c.}\)  \pm  u_+\bar{ u}_+\big]\\
%
& \pm \frac{\lambda}{2f}\frac{\dot\axion}{H}\big[ v_{\pm}\bar{ v}_{\pm}+i\( v_\pm \bar{ u}_\pm -\bar{ v}_\pm { u}_\pm\)  +  u_\pm \bar{ u}_\pm\big],
\end{align}
where we have integrated out the Gauss law constraint, and simplified things using the non-Abelian Coulomb gauge choice at eq.\ \eqref{eqn:vecgaugecon}. The action for the Higgs vector fluctuations and their interaction with the gauge fields is given by 
\begin{align}\nn
\frac{\delta^{2}\mathcal{L}_{\rm Higgs}}{k^2} = &  \frac{a^2}{2}\dHiggs^\pm{}'\bar{\dHiggs}^\pm{}' - \frac{a^2}{2} \dHiggs^\pm \bar{\dHiggs}^\pm  \mp a^2 \frac{m_\psi}{2x} \dHiggs^\pm \bar{\dHiggs}^\pm +a\frac{\mpsi }{x^2}( u^\pm\bar{\dHiggs}^\pm+\bar{ u}^\pm\dHiggs^\pm)\\ & -\frac{2a\mpsi (  u^\pm \bar{\dHiggs}^\pm{}' +\bar u^\pm  \dHiggs^\pm{}')}{x\(x^2\mp 2\mpsi x+2\mpsi^2+M_{\hvev}^2\)}-\frac{a^2   \dHiggs^\pm{}'   \bar{\dHiggs}^\pm{}'   }{2\(x^2\mp 2\mpsi x+2\mpsi^2+ M_{\hvev}^2\)},
\end{align}
Next, we need to eliminate the additional degree of freedom by imposing our gauge condition eq.\ (\ref{eqn:vecgaugecon}). We choose to eliminate $v^{\pm}$ in favour of $u^{\pm}$  and write
\begin{align}
v^{\pm} = i\(\frac{\mpsi}{x} \mp 1\) u_{\pm} = i \f_{\pm} u_{\pm}.
\end{align}
Note that there is nothing special about this choice, and we could equally well have eliminated $u^{\pm}$ in favour of $v^{\pm}$. We further canonically normalize the fields, introducing
\begin{align}
\v_{\pm} = & \sqrt{2(1+\f_{\pm})}u_{\pm}\\
\Xi_{\pm} = & a\sqrt{\frac{x^2\mp 2\mpsi x+2\mpsi^2 }{x^2\mp 2\mpsi x+2\mpsi^2+ M_{\hvev}^2}}i\dHiggs^{\pm} = a \g \xi^{\pm}.
\end{align}
Organizing the fields into the vector
\begin{align}
\Delta_V^{\pm} = (\v_{\pm}, \Xi_{\pm})
\end{align}
the action can be put into the form
\begin{align}
\delta^2\mathcal{L}_V = \frac{k^2}{2}\[\Delta^{\dagger}{}'_{V\pm}T_{\pm}\Delta'_{V\pm}+\Delta^{\dagger}{}'_{V\pm}K_{\pm}\Delta_{V\pm}+\Delta^{\dagger}{}_{V\pm}K^\dagger_{\pm}\Delta_{V\pm}'+\Delta^{\dagger}_{V\pm}\Omega^2_{\pm}\Delta_{V\pm} \], \quad T_{\pm} = \mathbb{1}
\end{align}
where the matrix $K_{\pm}$ is anti-Hermitian, while $\Omega^2_{\pm}$ is symmetric. These matrices have components
\begin{align}
K_{\pm, 21} = \mp\frac{2 M_{Z_0} \mpsi}{\left(2 \mpsi^2\mp2 \mpsi x+x^2\right) \sqrt{M_{Z_0}^2+2\mpsi^2+2
  \mpsi x+x^2}},
\end{align}
with $K_{11} = K_{22} = K_{12} = 0$, and
\begin{align}
\Omega^2_{\pm, 11} = & -1-\frac{2 \left(M^2+\mpsi^2+1\right)}{x^2} -\frac{\left(M^2+2\right)
   x}{2\mpsi \left(2 \mpsi^2\mp 2 \mpsi x+x^2\right)}\pm \frac{M^2+4
   \mpsi^2+2}{2\mpsi x}\\ \nn&+\frac{2 \mpsi^2 \left(M_{Z_0}^2-3 \left(2 \mpsi^2\mp 2 \mpsi
   x+x^2\right)\right)}{2\left(2 \mpsi^2\mp 2 \mpsi x+x^2\right)^2 \left(M^2+2 \mpsi^2\mp 2
   \mpsi x+x^2\right)}
\end{align}

\begin{align}
\Omega^2_{\pm, 22} = & -1+\frac{1}{x^2}\pm \frac{2 M_{Z_0}^2+M_{Z_0}^4+4 M_{Z_0}^2 \mpsi^2+4 \mpsi^4}{2 \mpsi \left(M_{Z_0}^2+2
   \mpsi^2\right) x}-\frac{\mpsi^2}{\left(2 \mpsi^2\mp 2 \mpsi
   x+x^2\right)^2}\\ \nn& +\frac{2 M_{Z_0}^2 \mpsi+4 \mpsi^3\mp (2 M_{Z_0}^2 x+M_{Z_0}^4 x)}{2 M_{Z_0}^2 \mpsi \left(2
   \mpsi^2\mp 2 \mpsi x+x^2\right)} -\frac{M_{Z_0}^2+\mpsi^2}{\left(M_{Z_0}^2+2 \mpsi^2\mp 2
   \mpsi x+x^2\right)^2}\\ \nn&+\frac{M_{Z_0}^4-4 M_{Z_0}^2 \mpsi^2-4 \mpsi^4+2 M_{Z_0}^2 \mpsi x}{M_{Z_0}^2
   \left(M_{Z_0}^2+2 \mpsi^2\right) \left(M_{Z_0}^2+2 \mpsi^2\mp 2 \mpsi x+x^2\right)}
\end{align}

\begin{align}
\Omega^2_{\pm, 12}  = &\frac{1}{\left(2 \mpsi^2\mp2 \mpsi x+x^2\right)^2 \sqrt{M_{Z_0}^2+2 \mpsi^2\mp2 \mpsi x+x^2}}\\ \nn& \times \Bigg( \mp\frac{2 \left(M_{Z_0}^3 \mpsi^3+2 M_{Z_0} \mpsi^5+2 M_{Z_0} \mpsi^3\right)}{x} \mp M_{Z_0} \mpsi x
   \left(M_{Z_0}^2+8 \mpsi^2+2\right) \\ \nn&+2 M_{Z_0} \mpsi^2 \left(M_{Z_0}^2+4 \mpsi^2+2\right)-\frac{2
   \left( \mpsi\mp x\right)M_{Z_0}^3 \mpsi }{M^2+2 \mpsi^2\mp2 \mpsi x+x^2}+4 M_{Z_0}
   \mpsi^2 x^2\mp M_{Z_0} \mpsi x^3\Bigg)
 \end{align}
and $\Omega^2_{\pm, 21} = \Omega^2_{\pm, 12}$. In the far past, the action quickly becomes diagonal (as $x^{2}$), and the fields become free. They are thus quantized in the usual way, as free plane waves.

In figure \ref{fig:vectormodes} we plot the evolution of the norm of the vector modes for the gauge fields modes, $v^{\pm}$, and the Goldstone, $\xi^{\pm}$. As is evident from the figure,  the addition of the Higgs leads to a temporary exponential instability in the vector modes that begins near $k\tau \sim M_{Z_0}$. At late times, the gauge vectors decay as $1/a$, while the Higgs vector freezes out.  These modes do not imprint any signatures on large scales due to the fact that their contribution to the vorticity and anisotropic stress is  suppressed by additional factors of the scale factor.

%%%%%%%%%%%%%%%%%%%%%%%%%%%%%%%%%%%%
%%%%%%%%%%%%%%%%%%%%%%%%%%%%%%%%%%%%
%%%%%%%%%%%%%%%%%%%%%%%%%%%%%%%%%%%%
\vskip 2cm

%\newpage
\bibliographystyle{JHEP}
\bibliography{HiggsedCNI}

%%%%%%%%%%%%%%%%%%%%%%%%%%%%%%%%%%%%
%%%%%%%%%%%%%%%%%%%%%%%%%%%%%%%%%%%%
%%%%%%%%%%%%%%%%%%%%%%%%%%%%%%%%%%%%

\end{document}